\documentclass[final,times,twocolumn,floatfix,prd,authoryear, nofootinbib,aps,eqsecnum,superscriptaddress,preprintnumbers,tightenlines]{revtex4}
\usepackage{graphicx,float}
\usepackage{bm}
\usepackage{newtxmath,newtxtext}
\usepackage{multirow}
\usepackage{epsfig}
\usepackage[dvipsnames]{xcolor}
\usepackage[colorlinks,linkcolor=blue,citecolor=red,urlcolor=blue]{hyperref}
\usepackage{makecell}

\newcommand{\be}{\begin{equation}}
	\newcommand{\ee}{\end{equation}}

\begin{document}
	
	\title{Nonminimally Coupled Quintessence with Double Exponential Potential: Observational Evidence Against Big Crunch Singularity}
	\author{Yerlan Myrzakulov}
	\email{ymyrzakulov@gmail.com}
	\affiliation{Department of General and Theoretical Physics, L. N. Gumilyov Eurasian National University, Astana 010008, Kazakhstan}
	
	\author{Saddam Hussain}
	\email{saddamh@zjut.edu.cn}
	\affiliation{Institute for Theoretical Physics and Cosmology, Zhejiang University of Technology, Hangzhou 310023, China}
	\author{Mohd Shahalam}
	\email{mohdshahamu@gmail.com}
	\affiliation{Department of Physics, Integral University, Lucknow 226026, India}
	
	\begin{abstract}
	We investigate a class of scalar field dark energy models non-minimally coupled to gravity, characterized by a double exponential potential and parameterized coupling $\xi$. We study the cosmological dynamics for a recently proposed descending dark energy model, namely, Q-SC-CDM. Initially, we choose distinct values of coupling parameter. For some values of $\xi$, the evolution of the universe is split up into three different phases: {\it decelerated expansion (early time), accelerated expansion (late-time) and slow-contraction (future era)}, and provide Big Crunch Singularity at distant future. In other scenario, the phase of slow-contraction vanishes, cosmic acceleration is obtained at current epoch, and the universe gets de-Sitter expansion at distant future. We confront the model with various datasets, including Cosmic Chronometers, Type Ia Supernovae (Pantheon+, DES, and Union 3), and Baryon Acoustic Oscillation measurements from DESI. Our analysis reveals that observational constraints naturally favor regions of parameter space in which the model avoids future singularities and slow-contraction phases, even for relatively small values of $\xi \simeq 0.12$. The preferred solutions yield values of $\Omega_{m}$, $H_0$, and $r_d$ that are consistent with those of $\Lambda$CDM at the $68\%$ confidence level. In contrast, the models with fixed couplings $\xi=0.3$ and $\xi=0.5$, which predict $H_0>70$ km/s/Mpc, are strongly disfavored relative to $\Lambda$CDM by the current datasets. Finally, the phase-space analysis confirms that the observationally constrained model with $\xi\simeq 0.12$ evolves toward a stable de Sitter attractor.
\end{abstract}
	
	\maketitle
	
	\tableofcontents
	
	\section{Introduction}
	\label{sec:intro}
	\noindent In recent years, scalar fields have become increasingly significant in cosmology and have an equation of state (EoS) parameter less than $-1/3$ to characterize the cosmic acceleration. They are used in various phenomenological models such as quintessence \cite{wet,ratra,sami06,sahni00}. In this case, a scalar field rolls down a potential energy function $V(\phi)$ that decreases monotonically over time. A wide range of different types of potential have been studied in the literature \cite{Peebles98,Turner97,Cald98,Zlatev,bamba,alam19,alam18,alam17}. In Planck mass units, the current value of $V(\phi)$ is minuscule. This indicates that the scalar field $\phi$ has a very low energy. There may be a potential that is currently shallow but will eventually reach more negative values. A specific configuration of $V(\phi)$ governs the dynamics of the scalar field and its influence on cosmic expansion. As the scalar field progresses, it can initiate various phases of the universe, ranging from accelerated expansion to gradual contraction. Recent studies have indicated that quintessence models featuring potentials that transition to negative values may lead to a future collapse of the universe, referred to as Big Crunch Singularity \cite{Kal03,Garri04,Peri05,Yun04,alam16}. Alternatively, the universe may either collapse or enter a sequence of cycles characterized by an initial expansion followed by a contraction. Cyclic cosmology posits that the universe undergoes an infinite number of cycles, each commencing with a `Big Bang' and concluding with a `Big Crunch'. Thus, the universe does not originate from a singularity; instead, it begins with a `Big Bang' that leads to a temporary expansion before contraction ensues. The fundamental concept of the cyclic model is that the universe can circumvent issues associated with the initial singularity by experiencing an endless series of cycles, each of which eliminates any anomalies that arose in the previous cycle. For instance, in the early universe, the scalar field may not dominate the energy density, leading to conventional cosmological evolution, such as radiation dominance followed by matter dominance. However, as the scalar field progresses down its potential, it may eventually come to dominate, resulting in accelerated expansion. These transitions from matter dominance to accelerated expansion can occur in a smooth and continuous manner, governed by the evolution of the scalar field. This ensures that the evolution of the universe remains steady and gradual, devoid of sudden or disruptive changes. As the scalar field descends the potential, the kinetic energy rises while the potential energy diminishes. Eventually, the kinetic energy surpasses the potential energy. At this juncture, the total energy density of the universe reaches zero, indicating that the Hubble parameter, which quantifies the rate of expansion of the universe, also becomes zero. Consequently, the expansion of the universe ceases entirely and begins to contract. The contraction phase is anticipated to be gradual, implying that the universe will continue to contract at a diminishing rate over an extended duration.
	
	In the standard quintessence scenario, the energy density of the scalar field minimally coupled to gravity mimics the effective cosmological constant. The detailed evolution is obviously dependent on a particular form of the potential, but the $\phi^2$ contribution can be regarded as a leading order term of expansion of the potential $V(\phi)$. We incorporate the non-minimal coupling constant to the quintessence scenario. Lykkas and Perivolaropoulos demonstrate that certain values of the non-minimal coupling parameter can prevent the cosmic apocalypse singularity in scalar-tensor quintessence with a linear potential \cite{Peri15}. {Additionally, in Ref.~\cite{Andriot:2024jsh}, the authors explicitly demonstrated that a quintessence field with an exponential potential yields an eternal accelerating solution in the future.} For many years, researchers have examined the scalar tensor theories, which have a scalar field that is non-minimally coupled (NMC) to gravity. In order to reconcile Mach's principle with general relativity, the first well-known instance of non-minimal coupling \`a $l$a Brans-Dicke theory was put forth in 1961 \cite{bd}. According to this theory, the gravitational constant is replaced by a scalar field $\phi$ entering the action in a particular combination with a Riemannian curvature as $\phi^2 R$. Other types of scalar-tensor action were studied after the Brans-Dicke proposal: a well-known example of a NMC scalar field system is given by $F(\phi)R$ coupling with $F(\phi)=1- \kappa^2 \xi \phi^2$. Such a theory has quite rich cosmic dynamics that need consideration. For a recent breakthrough, a NMC Higgs field because of a large coupling $\xi$ could result in a successful inflation, which is otherwise impossible \cite{bezru}. The construction of a dark energy model is very interested in NMC scalar field systems because of their unique characteristics \cite{a1,a2,a3,a4,a5,a6,a7,a8,a9,a01,a02,a03,a04,a05,a06,a07,a08,a09,a010,a011,a012,a013,a014,a015,a016,sunny,25a1,25a2,25a3,25a4}. Non-minimal coupling, for example, may enable phantom crossover and result in cosmic scaling solutions that are relevant to dark energy models. The general characteristics of a NMC system with $F(\phi)=1- \kappa^2 \xi \phi^2$ are phantom scaling solutions \cite{a10}. In the literature, two values of the coupling constant are frequently considered: $\xi=0$ (minimal coupling) and $\xi=1/6$ (conformal coupling). However, in many cases, $|\xi|\gg1$ (strong coupling) has also been used \cite{x1,x2,x3,x4,x5,x6}. Recently, the value of $\xi$ was estimated from a distant supernova \cite{xs}.
	
	Dynamical system theory has been widely applied in cosmology to provide a broad understanding of dynamics for a variety of cosmological models. This approach has the benefit of having some sort of ``machines" that uses a straightforward programmed algorithm to derive asymptotic solutions. In order to rewrite the original system as a system of first-order differential equations, a new set of variables must be introduced. This approach allows for the derivation of asymptotic solutions, and stability can be assessed using a straightforward programmed algorithm. Traditionally, physical cosmological solutions are derived from the combination of phase portraits and the stability of critical points. The ability to display any system trajectory that is permissible for every initial condition is the primary benefit of dynamical system analysis. Consequently, generic pathways to the accelerating phase (de-Sitter attractor) can be categorized. This paper aims to explore the cosmological dynamics of the Quintessence-Driven Slow-Contraction Cold Dark Matter (Q-SC-CDM) model and to examine its stationary points and stability. The Q-SC-CDM model was initially introduced by C. Andrei {\it et al.} in the framework of descending dark energy and the end of cosmic expansion \cite{stein,alam25}. In this research, we examine the cosmological dynamics of a NMC scalar field model with double exponential potential and a specific form of $F(\phi)$. We limit ourselves to the functional form $F(\phi) = 1 - \kappa^2 \xi \phi^2$. We shall analyze the cosmological dynamics of the model under consideration and discuss the regions of deceleration, cosmic acceleration and slow-contraction. Furthermore, we shall conduct a phase space analysis of the Q-SC-CDM model, and investigate the existence of a stable late-time de-Sitter attractor solution that corresponds to an EoS of $-1$ and an energy density parameter $\Omega_{\phi} = 1$. 
	
	{The paper is structured as follows. In Section~\ref{sec:EOM}, we set up the theoretical model and use numerical analysis to identify which trajectories are cosmologically viable and what the future asymptotics are as functions of the model parameters (Section~\ref{sec:CD}). In Sections~\ref{sec:EDA}--\ref{sec:RES}, we then use observational data to determine whether current data prefer the model parameters to lie within the viable region identified from the first analysis. In Section~\ref{sec:phase}, we construct the autonomous system for dynamical systems analysis, linking the phase-space structure to the observational constraints obtained in the preceding sections. Finally, we summarize our findings in Section~\ref{sec:conc}.}
	
	\section{Equations of motion}
	\label{sec:EOM}
	\noindent A NMC scalar field model refers to a theoretical framework in which a scalar field is coupled to gravity not only through the usual minimal kinetic and potential terms, but also directly to the Ricci scalar $R$ of spacetime. This coupling modifies the Einstein field equations and allows for richer dynamics in the evolution of the universe.
	The action for such a model typically takes the form \cite{Faraoni:2000wk,orest1,orest2,Sami:2012uh,Shahalam:2019jgs,Shahalam:2020lcc}.
	\be
	\label{eq:Lagrangian}
	S=\frac{1}{2}\int{\sqrt{-g}d^4x\Big{[} \frac{1}{\kappa^2}
		F(\phi) R-\Big{(}g^{\mu\nu}\phi_{\mu}\phi_{\nu}+2V(\phi)\Big{)}\Big{]}}+S_M \ee
	{ Here, $F(\phi)=1-\kappa^2 \xi \phi^2$, where $F(\phi)$ denotes a scalar field dependent function that characterizes the nonminimal coupling.} The $\kappa^2=8 \pi G = 1/M_{Pl}^2$, $S_M$ represents the action of matter, $V(\phi)$ denotes the potential of the scalar field, and $g$ is the determinant of the metric tensor $g_{\mu \nu}$. Varying the action (\ref{eq:Lagrangian}) yields the equations of motion in a spatially flat FLRW background, which are given by
	\be \label{eq:Friedphi} H^2=\frac{\kappa^2}{3}\left(\frac{1}{2}{\dot
		{\phi}}^{2}+V(\phi)+6\xi H \phi \dot {\phi}+3 \xi H^{2} \phi^2+\rho_m
	\right), \ee
	\begin{multline}
		\label{eq:Friedphi2} R=\kappa^2 \bigg(-{\dot {\phi}}^{2} +4
		V(\phi)+18\xi H \phi \dot{\phi} +\xi R \phi^2  +6 \xi {\dot
			{\phi}}^{2}+\\ 6 \xi \phi \ddot {\phi}+\rho_m(1-3 \omega_{m}) \bigg),
	\end{multline}
	\begin{multline}
		\frac{\ddot a}{a} = - \frac{\kappa^2}{3} \bigg((1-3 \xi) \dot{\phi}^2-V(\phi)-3\xi H \phi \dot{\phi}-3\xi \phi \ddot {\phi}-\\ 3\xi \left(\frac{\ddot a}{a} \right) \phi^2+ \frac{\rho_{0m}}{2a^3}\,\bigg),
		\label{eq:add}
	\end{multline}
	
	\begin{eqnarray}
		\label{eq:KGphi}
		&&\ddot {\phi}+3 H \dot {\phi}+\xi R \phi +V'(\phi)=0.
	\end{eqnarray}
	Where Ricci scalar $R=6(2H^2+\dot{H})$. The pressure and energy density of matter are denoted by $p_m$ and $\rho_m$, and are related as $p_m=\omega_m \rho_m$; the $\omega_m$ represents the EoS for matter. The energy density and pressure for a NMC scalar field model are defined as
	\begin{eqnarray}
		\label{eq:rhophi}
		\rho_{\phi}&=&\frac{1}{2}{\dot {\phi}}^{2}+V(\phi)+6\xi H \phi \dot {\phi}+3 \xi H^{2} \phi^2,\\
		p_{\phi}&=&\frac{1}{2}{\dot {\phi}}^{2}-V(\phi)-\xi \phi^2 \bigg(4 H \phi \dot
		{\phi} +\nonumber \\&& 2\dot {\phi}^{2} +2 \phi \ddot {\phi} +(2\dot{H}+3H^{2})  \bigg)\ .
	\end{eqnarray}
	From equation (\ref{eq:Friedphi}), we get following constraint equation.
	\begin{eqnarray}
		\frac{\kappa^2 \rho_{\phi}}{3H^2}+\frac{\kappa^2 \rho_{m}}{3H^2}&=&1\\
		\Omega_{\phi}+\Omega_{m}&=&1
	\end{eqnarray}
	where $\Omega_{\phi}=\frac{\kappa^2 \rho_{\phi}}{3H^2}$ and $\Omega_{m}=\frac{\kappa^2 \rho_{m}}{3H^2}$ represent the energy density parameter for scalar field and matter, respectively. The EoS $w_{\phi}$ and total EoS $w_{\rm eff}$ for a NMC scalar field model are given as
	\begin{eqnarray}
		w_{\rm eff}&=&-1-\frac{2 \dot{H}}{3H^2}\\
		w_{\phi} &=& \frac{w_{\rm eff}-w_m \Omega_m}{1-\Omega_m}    
	\end{eqnarray}
	\begin{figure}
		\includegraphics[scale=0.5]{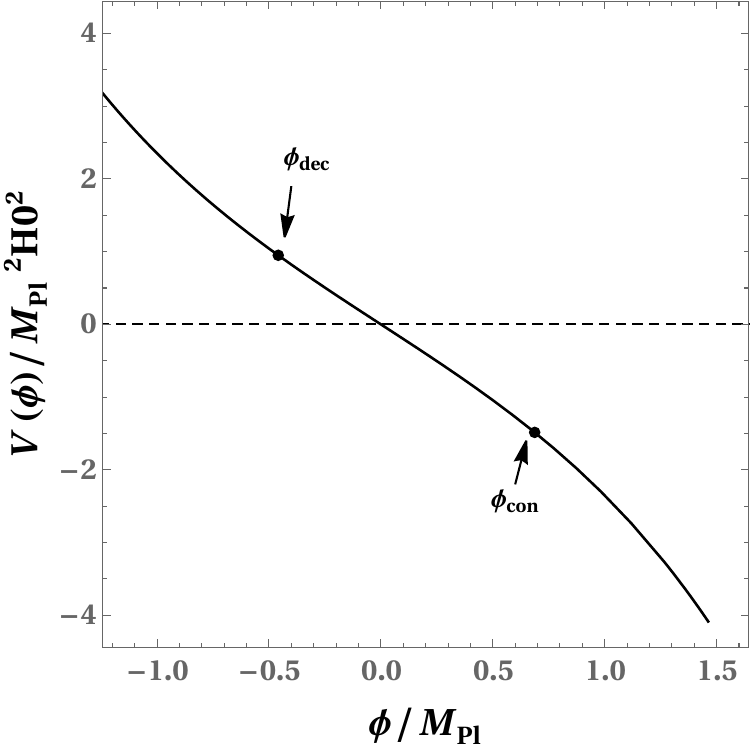}
		\caption{\small The figure shows the evolution of potential (\ref{eq:pot}) vs field. Initially, the potential is positive for negative values of scalar field. As field evolves, the universe expands with cosmic acceleration and remains so until $\phi_{dec}$. On further evolution of field, the potential becomes more negative. To this effect, the expansion factor $a(t)$ collapses at $\phi_{con}$, and correspondingly Hubble parameter $H(t)$ becomes zero, see Fig. \ref{fig:NMC1}.}
		\label{fig:pot}
	\end{figure}
	\noindent In the following subsection, we shall study the cosmological dynamics of descending dark energy model with non-minimal coupling.

	\subsection{Cosmological Dynamics of Descending Dark Energy}
	\label{sec:CD}
	
	The descending dark energy model suggests that dark energy is not a constant but gradually loses energy over time. In such a model, the energy density of dark energy decreases as the universe evolves, either through a direct decay into other forms of energy or as a function of the scale factor of the universe. This is different from the standard cosmological model, which assumes that dark energy remains constant in its density (described by the cosmological constant $\Lambda$). Recently, a new model of descending dark energy, namely, Q-SC-CDM was proposed by C. Andrei $et$ $al.$ \cite{stein}. The Q-SC-CDM scenario proposes that the universe underwent a period of slow-contraction rather than acceleration. In this case, quintessence acts as the driving force for the contraction and evolves in such a way that it causes the cosmic scale factor to shrink over time. We use the following potential proposed by C. Andrei $et$ $al.$ \cite{stein,alam25}.
	\begin{equation}
		V(\phi)=V_0 e^{-\kappa \alpha \phi}-V_1 e^{\kappa \beta \phi},
		\label{eq:pot}
	\end{equation}
	where $V_0$ and $V_1$ $(\alpha $ and $\beta)$ are constants of mass dimension four (zero). To study the cosmological behavior for NMC scalar field model, we use following dimensionless parameters 
	\begin{eqnarray}
		H_{0}t &\longrightarrow &t,  \notag \\
		\kappa \phi  &\longrightarrow &\phi,  \notag\\
		\frac{\kappa^2 V_{0}}{H_{0}^{2}} &\longrightarrow &V_{0}.  \notag\\
		\frac{\kappa^2 V_{1}}{H_{0}^{2}} &\longrightarrow &V_{1}. 
		\label{eq:dimension}
	\end{eqnarray}
	The equations (\ref{eq:add}) and (\ref{eq:KGphi}) in re-scaled form can be written as
	\begin{multline}
		\frac{\ddot{a}}{a}=\frac{1}{1-\xi \phi^2} \left( -\frac{(1-3\xi)\dot{\phi}^{2}}{3} +\frac{V_0 e^{-\alpha \phi}- V_1 e^{\beta \phi}}{3}+\xi \frac{\dot{a}}{a} \phi \dot{\phi}\right.\\ \left. +\xi \phi \ddot{\phi} -\frac{\Omega _{0m}}{2a^{3}} \right),
		\label{eq:add1}
	\end{multline}
	\begin{equation}
		\ddot{\phi}+3\frac{\dot{a}}{a} \dot{\phi}+6\xi \left( \frac{\dot{a}^2}{a^2}+\frac{\ddot{a}}{a} \right)\phi-\alpha V_0 e^{-\alpha \phi}- \beta V_1 e^{\beta \phi} =0. 
		\label{eq:phidd1}
	\end{equation}

	{\begin{figure}[tbh]
			\resizebox{\columnwidth}{!}{\includegraphics{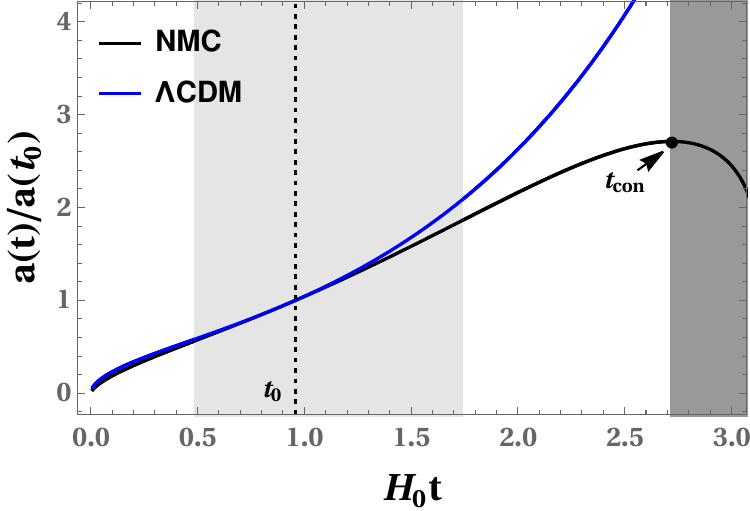}\includegraphics{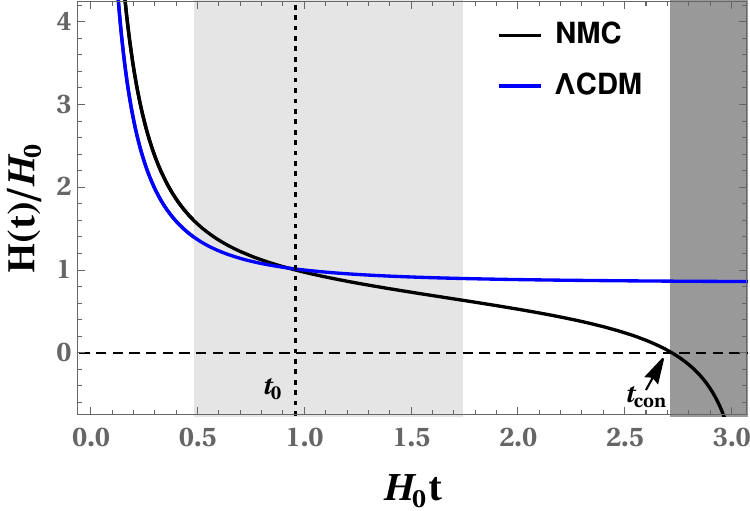} \includegraphics{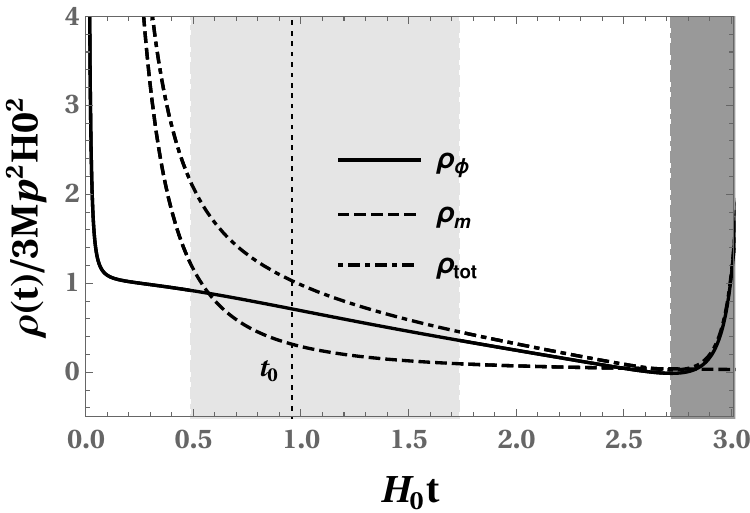}
			}
			\resizebox{\columnwidth}{!}{\includegraphics{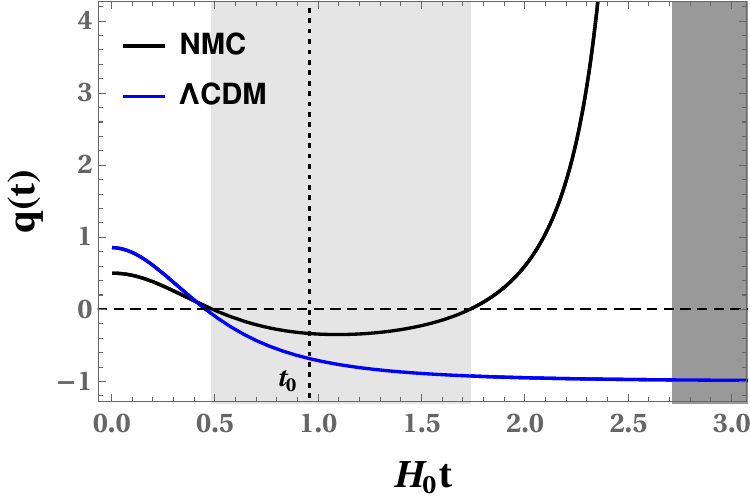} \includegraphics{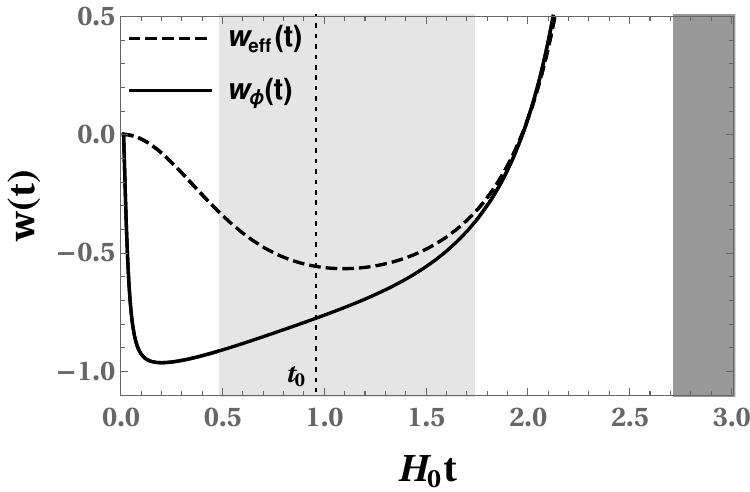}
				\includegraphics{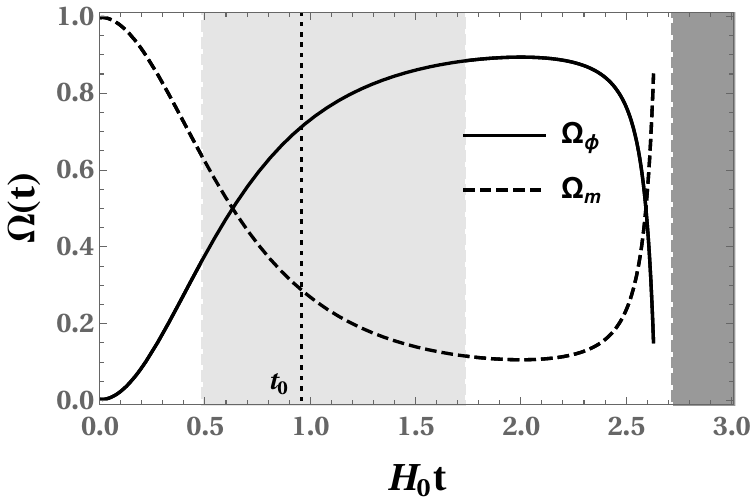}
			}
			\caption{The numerical evolution of expansion factor $a(t)$, Hubble parameter $H(t)$, energy density $\rho(t)$, deceleration parameter $q(t)$, equation of state $w(t)$ and density parameter $\Omega(t)$ versus $H_0t$ for the NMC model with $\xi=0.1$, $\alpha=\beta=1$, $V_0=3.01$, $V_1=0.33$ and $\Omega_{0m} = 0.31$.  The Unshaded, light gray shaded and dark shaded regions represent the periods of decelerated expansion, accelerated expansion $(H > 0)$ and slow contraction $(H < 0$ at $t = t_{\rm con})$, respectively.}
			\label{fig:NMC1}
		\end{figure}

		At early times, the universe was matter dominated and the field $\phi$ was frozen due to large Hubble damping. Therefore, it is straight forward to solve equations (\ref{eq:add1}) and (\ref{eq:phidd1}) numerically with the following initial conditions $(t \rightarrow t_i \simeq 0)$:
		\begin{eqnarray}
			a(t_{i}) &=&\left( \frac{9\Omega _{0m}}{4}\right) ^{1/3}t_{i}^{2/3},  \notag
			\\
			\dot{a}(t_{i}) &=& \frac{2}{3}\left( \frac{9\Omega _{0m}}{4}\right)
			^{1/3}t_{i}^{-1/3},  \notag \\
			\phi (t_{i}) &=&\phi _{i},  \notag \\
			\dot{\phi}(t_{i}) &=&0.  \label{eq:init}
		\end{eqnarray}
		We consider a prior value of $\Omega_{0m} = 0.31$. By tuning $\phi_i$, we get following parameters at the present epoch ($t_{0}$).
		\begin{eqnarray}
			a(t_{0}) &=& 1,  \notag \\
			H(t_{0}) &=& 1. 
			\label{eq:parm}
		\end{eqnarray}
		where, $t_{0}$ is present time at which scale factor is unity. In Fig. \ref{fig:pot}, we exhibit the evolution of the potential (\ref{eq:pot}) versus field $\phi$. Initially, potential shows positive behavior for negative scalar field, as time passes, the universe gets expansion with cosmic acceleration, and continues to evolve until $\phi_{\rm dec}$ at which accelerated expansion turns decelerated expansion, on further evolution of $\phi$, the potential becomes more negative, and the scale factor $a(t)$ collapses ($t=t_{\rm con}$) correspondingly Hubble parameter $H(t)$ goes to zero, see upper panels of Fig. \ref{fig:NMC1}. This implies that the expansion phase ends and slow-contraction begins. The lower panels of Fig. \ref{fig:NMC1} display the evolution of the deceleration parameter $q(t)$, EoS $w(t)$ and energy density parameter $\Omega(t)$ versus cosmic time. Under the chosen model parameters with $\xi=0.1$, we get $q=-0.33$, $w_{\rm eff}=-0.56$ and $w_{\phi}=-0.78$ at the present epoch. {On the other hand for $\xi=0.5$ (rest of the parameters are same as mentioned in the caption of Fig. \ref{fig:NMC1}), we find $q=-0.53$, $w_{\rm eff}=-0.72$, $w_{\phi}=-0.98$ and $\Omega_{\phi}=0.70$ at current epoch, and the numerical evolution of cosmological parameters are shown in Fig. \ref{fig:NMC2}.} 
		
		\begin{figure}[tbh]
			\resizebox{\columnwidth}{!}{\includegraphics{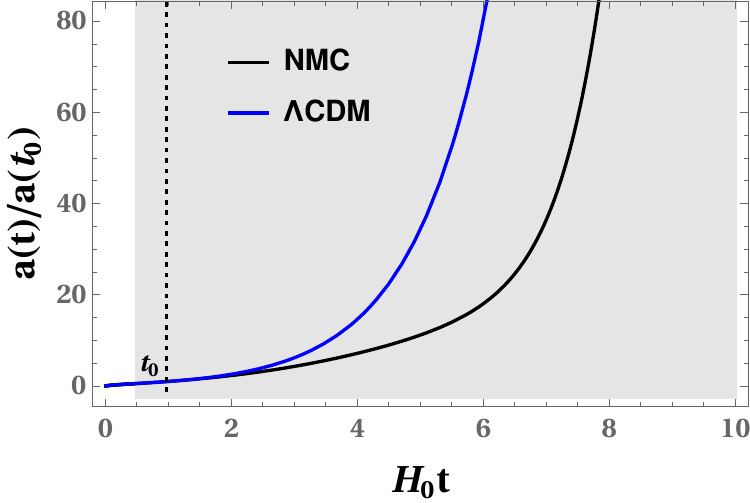} \includegraphics{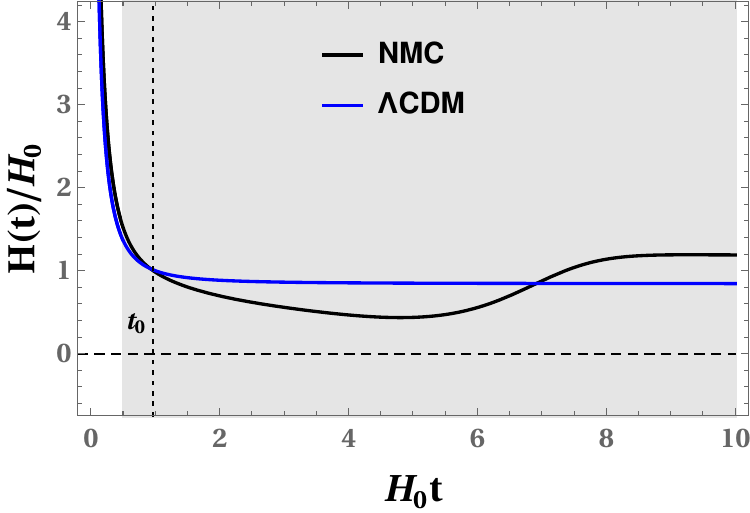}
				\includegraphics{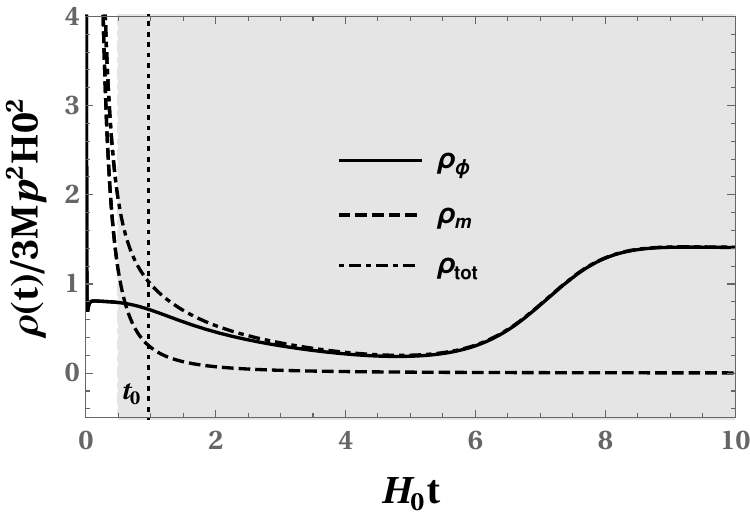}}
			\resizebox{\columnwidth}{!}{\includegraphics{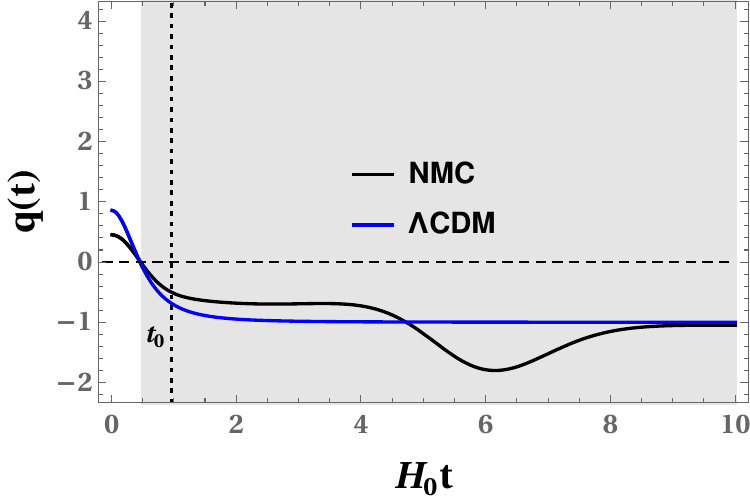}
				\includegraphics{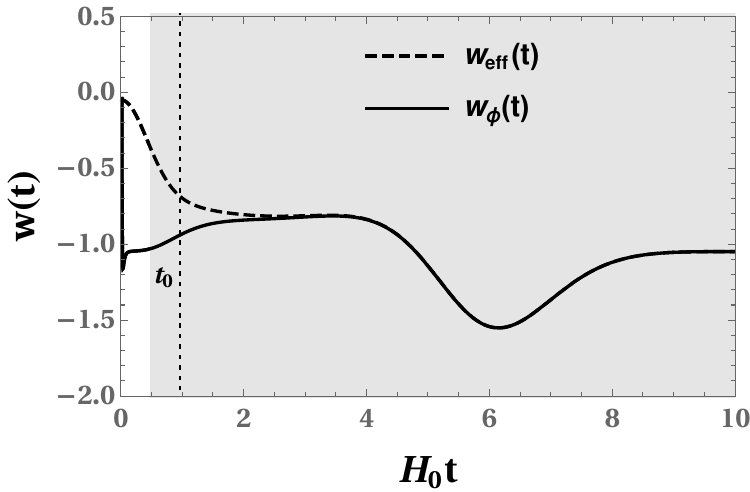}
				\includegraphics{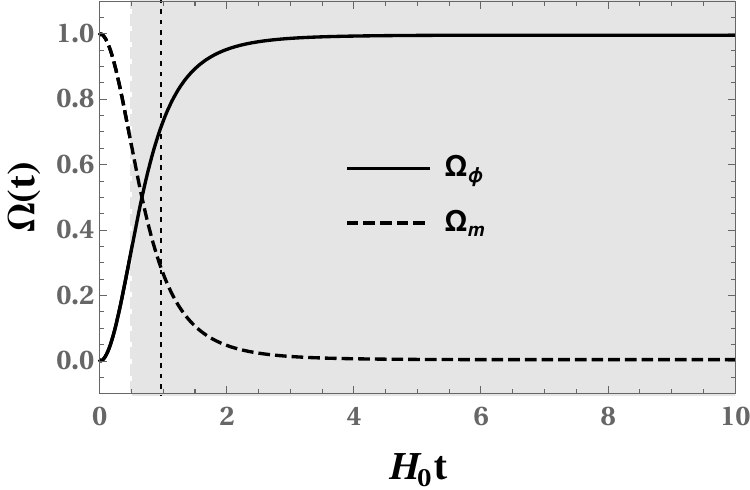}}
			
			\caption{The figure is similar to Fig. \ref{fig:NMC1} except $\xi=0.5$. In this case, we get $q=-0.53$, $w_{\rm eff}=-0.72$, $w_{\phi}=-0.98$ and $\Omega_{\phi}=0.70$ at current epoch. Moreover, the phase of slow-contraction vanishes, and we obtain cosmic acceleration at current epoch and de-Sitter expansion in distant future.}
			\label{fig:NMC2}
		\end{figure}
		
		Let us make some comments on the value of non-minimal coupling $\xi$. As $\xi \rightarrow 0$, the standard quintessence scenario is recovered. In this case, the evolution of the universe is divided into three different phases such as deceleration (past era), acceleration (current epoch) and slow-contraction (future era) \cite{stein}\footnote{{It is important to note that, for a wide range of model parameters $(\alpha,\beta)$, the models that exhibit a future slow-contraction phase fail to produce a sufficiently prolonged matter-dominated era in the past, which is essential for structure formation and for describing the observed Universe (see Figs.~\ref{fig:testing_different_xi_2} and \ref{fig:xi_0_alpha_beta}). Consequently, despite admitting a slow-contraction epoch, such models cannot be regarded as viable alternatives to $\Lambda$CDM.}}. In quintessence models where potentials become negative may give rise to collapse of the universe in future, which is known as Big Crunch Singularity. Such singularity can be avoided in scalar-tensor theories with non-minimal coupling \cite{Peri15}. 
		
		{To investigate this effect, we consider different values of the coupling parameter in the range $0.01<\xi<1$ and numerically evolve the system of equations from Eq.~\eqref{x_prime}--Eq.~\eqref{om_prime} while varying the model parameters within $0.01<\alpha<2$ and $0.1<\beta<2$. The resulting evolution of the normalized Hubble parameter, $h \equiv H/H_0$, and the effective equation of state, $w_{\rm eff}$, is presented in Appendix~\ref{appen:variation_of_xi_with_alpha_beta}. From Fig.~\ref{fig:xi_1_alpha_beta}, we find that for $\xi<0.2$ a slow-contraction phase can occur when $(\alpha,\beta)$ take sufficiently large values. Nevertheless, for some parameter combinations satisfying $-1.5 < \log_{10}(\alpha \cdot \beta) < -0.5$, the model instead approaches a future state with nearly constant $h$. In contrast, when $\xi>0.2$, none of the trajectories corresponding to the explored combinations of $\alpha$ and $\beta$ exhibits a slow-contraction phase with $w_{\rm eff}>-1/3$ in the future. Therefore, a sufficiently large non-minimal coupling plays a crucial role in avoiding the Big Crunch singularity over the considered parameter space.
			
			However, as discussed above and further demonstrated in the next section, parameter regions associated with future slow-contraction behavior fail to produce a sufficiently prolonged matter-dominated epoch. Consequently, such solutions cannot be regarded as physically viable alternatives to $\Lambda$CDM. Furthermore, the non-minimal coupling induces a transient variation in the Hubble parameter before the Universe eventually settles into a stable de Sitter phase. This behavior distinguishes the model from a pure cosmological constant scenario. Confronting the model with cosmological observations therefore provides a direct test of whether such deviations can be detected with the precision of current datasets.} 		

		\section{Equations for Data analysis}
		\label{sec:EDA}
		
		{In the previous section, we evolved the field--fluid variables with respect to cosmic time and showed the effect of the non-minimal coupling parameter on the system. The dynamics were obtained by considering specific values of the free parameters corresponding to the potential. The goal of the present section is to test the model against updated observational datasets and to obtain constraints on the model parameters. To achieve this, we reparameterize the dynamical variables into dimensionless form, such that the system of second-order differential equations can be mapped into first-order differential equations and thus can be directly implemented in a Python pipeline. We then evolve the dynamical variables with respect to the redefined time variable $(N = \ln a) = -\ln(1+z)$, where $z$ is the redshift, which facilitates obtaining the Hubble parameter $H(z)$ as a function of redshift. Consequently, geometric distances such as luminosity distances and angular diameter distances can be readily computed and compared with Type Ia Supernova observations. 
			
			The choice of redefined dimensionless dynamical variables is closely aligned with the dynamical system analysis framework, which is also applied to the present system in the next section. However, the reader should note that one can directly implement the dynamical system framework to obtain constraints on the model parameters \cite{Hussain:2024qrd,Hussain:2024jdt,Hussain:2024yee}. Nevertheless, for nonlinear systems, the autonomous system of equations may contain critical points where the system diverges. Such divergences lead to numerical instabilities, making it infeasible to obtain solutions for the dynamical variables across redshift. 
			
			To address stability, these divergences are often removed by redefining the time variable $(N \to \tilde{N})$ \cite{Bahamonde:2017ize}. However, such a transformation cannot be implemented when confronting the model with observations, since redefining time alters the redshift $(z \to \tilde{z})$. Therefore, for the present system, our choice of dynamical variables is:} 
		\begin{equation}
			x = \frac{\kappa \dot{\phi}}{\sqrt{6} H_0}, \quad \tilde{y} = \frac{\kappa \sqrt{V}}{\sqrt{3} H_0}, \quad \tilde{\Omega}_{m} =\frac{\kappa \rho_{m}}{3 H_0^2}, \quad  h = \frac{H}{H_0}, \quad z = \kappa \phi, \label{dimensionless_variable_data}
		\end{equation}
		where the field and fluid variables are normalized using the Hubble constant $H_0$. In terms of these variables, the Friedmann equation becomes: 
		\begin{equation}
			h^2 = x^2 + \tilde{y}^2 + 2 \sqrt{6} \xi x z h + \xi h^2 z^2 + \tilde{\Omega}_{m} \ .
			\label{hubble_new_eq}
		\end{equation}
		Solving the quadratic equation for $h$, we obtain the following valid root:
		\begin{equation}
			h = -\frac{\frac{1}{2} \sqrt{24 \xi ^2 x^2 z^2-4 \left(\xi  z^2-1\right) \left(-A^2+\tilde{\Omega}_{m}+x^2+y^2\right)}+\sqrt{6} \xi  x z}{\xi  z^2-1}\ ,
		\end{equation}
		which ensures a positive value of $h$ throughout the evolution. Since $h$ is a monotonically increasing positive function of redshift, we discard the other root, which yields unphysical negative values of $h$. In this study, we adopt a dual-exponential form of the potential, characterized by two parameters:
		\begin{equation}
			V(\phi) = V_{0} e^{-\kappa \alpha \phi}-V_{1} e^{\kappa \beta \phi} \ , 
		\end{equation}
		which results in $\tilde{y}$
		\begin{equation}
			\tilde{y}^2 = y^2 - A^2\ ,
		\end{equation}
		where 
		\begin{equation}
			y = \sqrt{\frac{\kappa^2 V_{0} e^{-\kappa \alpha \phi}}{3 H_0^2}}, \quad A = \sqrt{\frac{\kappa^2 V_{1} e^{\kappa \beta \phi} }{3 H_0^2}} \ . \label{y_def}
		\end{equation}
		{In the previous section, we redefined the dimensionless quantities $(V_0,V_1)$ in Eq.~\eqref{eq:dimension}, corresponding to $(\kappa^2 V_0/H_0^2)$ and $(\kappa^2 V_1/H_0^2)$, respectively. Here, these constants are absorbed into the variables $y$ and $A$ together with the exponential functions of $\phi$, which remain positive for any value of $\kappa\phi$. Consequently, in the present analysis it is not necessary to assign specific values to these constants, as was done in the evaluation of Figs.~\ref{fig:NMC1} and \ref{fig:NMC2}. Therefore, the quantity that primarily determines the overall dynamics and the occurrence of the slow-contraction phase is $\tilde{y}$, which corresponds to the effective potential.
			
			As shown in Fig.~\ref{fig:pot}, the effective potential can become negative when $A>y$, leading to a slow-contraction phase in which $H\rightarrow 0$. This behavior disappears whenever $y>A$. Therefore, in this and the following section, we focus on tracking the evolution of the variables $(y,A)$ rather than the specific values of the parameters $(V_0,V_1)$. Interested readers may recover the values of $(V_0,V_1)$ at the present epoch, $N=0$, from the corresponding values of $y(N=0)=y_0$, $z(N=0)=z_0$, and $A(N=0)=A_0$ obtained for a given choice of $\alpha$ and $\beta$. Hence, $V_0$ = $\frac{3 y_0^2 }{e^{-\alpha z_0}}$ and $V_1 = \frac{3 A_0^2}{e^{\beta z_0}}$.  }
		
		With these dynamical variables, the Ricci scalar can be expressed as: 
		\begin{multline}
			R = \frac{H_0^2}{1 - \xi z^2} \bigg(-6x^2 + 12 \tilde{y}^2 + 18 \sqrt{6} \xi h z x + 36 \xi x^2 + \\  6 \xi z \frac{\kappa \ddot{\phi}}{H_0^2} + 3\tilde{\Omega}_{m}(1-3w_m)\bigg) \ .
		\end{multline}
		For a flat FLRW metric, the Ricci scalar takes the form
		\begin{equation}
			R = 12 H^2 + 6 \dot{H}\ ,
		\end{equation}
		which allows us to express the time derivative of the Hubble parameter $\dot{H}$ by using the field equations, and given as:
		
		\begin{multline}
			\dfrac{\dot{H}}{H^2} = -2 + \frac{1}{6h^2(1-\xi z^2) \left(1+ \dfrac{6z^2 \xi^2}{1-z^2 \xi}\right)}\times\\ \bigg(-6x^2 + 12 \tilde{y}^2 + 36 \xi x^2 + 18\xi z \alpha (A^2 + y^2) + 3\tilde{\Omega}_{m}(1-w_m)\bigg) \ .
			\label{eq:Hdot}
		\end{multline}
		
		Analogous to the procedure in the preceding section, we derive the first-order autonomous system of equations from the field equations as: 
		\begin{eqnarray}
			x ' &= & \frac{1}{\sqrt{6}} \left(\frac{1}{1+ \frac{6 \xi^2 z^2}{1- \xi z^2}}\right) \bigg(-3\sqrt{6} x + \frac{3 y^2 \alpha}{h} + \frac{3 A^2 \beta}{h} - \nonumber \\ 
			&&\frac{\xi z}{h(1-\xi z^2) } \bigg[-6x^2+12\tilde{y}^2 + 18\sqrt{6} z \xi h x  +\\ \nonumber&& 36 \xi x^2  + 3 \tilde{\Omega}_{m}(1-3\tilde{\Omega}_{m})\bigg] \bigg),\label{x_prime}\\
			y' & = &  \frac{-1}{2h} y \sqrt{6} x \alpha, \label{y_prime}\\
			A'  & = & \frac{1}{2 h} A \beta \sqrt{6} x , \label{A_prime}\\
			z' & =& \frac{x \sqrt{6}}{h}, \label{z_prime}\\
			\tilde{\Omega}_{m}' & = & - 3 \tilde{\Omega}_{m} (1+ w_m) \label{om_prime} \, ,
		\end{eqnarray}
		where \(()' = d()/dN\), $N$ representing the number of e-folds \(N = \log a\).
		
		As we evolve the system toward higher redshifts, i.e., using the variable $N = -\ln(1+z)$,\footnote{Here, $z$ denotes the redshift. This should not be confused with the dynamical variable $z = \kappa \phi$. From this point onward, whenever the term redshift is used, it will be denoted by $z$.} we set initial conditions at the current epoch, corresponding to $N = 0$. It immediately follows from the Hubble Eq.~\eqref{hubble_new_eq} that the condition $h(0) = 1$ must be satisfied. As a result, the initial value of one of the dynamical variables—say $y$—can be expressed in terms of the others as:
		\begin{equation}
			y(0) = \sqrt{1 - x_0^2 + A_0^2 -2 \sqrt{6} \xi x_0 z_0 - \xi z_0^2 - \tilde{\Omega}_{0m}} \ .
		\end{equation} 
		Thus, although the system consists of five equations, the Hubble constraint Eq.~\eqref{hubble_new_eq} effectively reduces the parameter space by one degree of freedom. In the next section, we will iterate over the remaining parameters to obtain the best-fit values. Additionally, we define the cosmological parameters, such as the fractional energy densities as follows:
		
		\begin{multline}
			\Omega_{\phi} = \frac{\kappa^2 \rho_{\phi}}{3 H^2} = \frac{x^2 + \tilde{y}^2 + 2 \sqrt{6} \xi x z h + \xi h^2 z^2}{h^2}, \\ \Omega_{\rm m} = \frac{\kappa^2 \rho_{m}}{3 H^2}= \frac{\tilde{\Omega}_{m}}{h^2}, \quad \Omega_{\rm NCV} =  \dfrac{2 \sqrt{6} \xi x z h + \xi h^2 z^2 }{h^2} \ .
		\end{multline}
		
		Here, we adopt the minimally coupled analogy for the scalar field, where the energy density takes the standard form $\rho_{\phi} = \frac{1}{2}\dot{\phi}^2 + V(\phi)$. { Due to the non-minimal coupling, the field density now receives a contribution from $\xi$, where the non-minimal coupling density is identified as $\Omega_{\rm NCV}$. In general, the non-minimal coupling density does not need to be strictly positive, since it does not correspond to a physical quantity. However, the field density must be strictly positive, $0 \le \Omega_{\phi} = 1 - \Omega_{\rm m} \le 1$, and must remain within this range. Similarly, we define the equation of state corresponding to the scalar field as $\omega_{\phi} = p_{\phi}/\rho_{\phi}$, and the effective equation of state, which can be written as:
			\begin{equation}
				\omega_{\rm eff} = -1-\frac{2 \dot{H}}{3H^2}, \quad \omega_{\phi} = \frac{\omega_{\rm eff}}{\Omega_{\rm \phi}}\ ,
			\end{equation}
			where $\frac{\dot{H}}{H^2}$ is given by Eq. (\ref{eq:Hdot}). To illustrate the evolution across redshifts, we plot the cosmological parameters and dynamical variables for various values of $\xi$ in Figs. \ref{fig:testing_different_xi_1} and \ref{fig:testing_different_xi_2}. {Here, we have presented the behavior only for selected values of the model parameters; however, the qualitative features remain unchanged within their neighboring regions. A more detailed analysis is provided in Appendix~\ref{appen:variation_of_xi_with_alpha_beta}, where the model parameters are varied over a wide range. Guided by the behavior shown in Figs.~\ref{fig:xi_0_alpha_beta} and \ref{fig:xi_1_alpha_beta}, we select representative initial conditions and investigate the corresponding evolution of the cosmological and dynamical variables in detail. These results allow us to infer the effective behavior of the model. Furthermore, the numerical analysis provides insight into the range of behaviors admitted by the model and helps motivate the prior ranges that will later be adopted in the observational analysis. The parameter constraints obtained in the next section will ultimately determine the observational viability of the model.} We observe that for lower values ($\xi < 0.2$), the matter density remains bounded between 0 and 1 for certain parameter ranges; however, the field density either becomes negative or exceeds unity at higher redshifts, which is physically non-viable. A similar behavior is noted for higher values ($\xi \ge 0.3$), where the field density is initially positive and bounded within $[0, 1]$ for specific parameter spaces and initial conditions, but eventually becomes negative for $\alpha, \beta > 0$. 
			
			For $\alpha < 0$, the model with $\xi = 0.3$ produces a negative field energy density, while for $\xi = 0.1$, the field density diverges abruptly around $N \sim -2$. Additionally, the dynamical variables $y$ and $A$ become negative in the past for $\xi = 0.1$, contradicting the definition in (\ref{y_def}). Consequently, such parameter ranges are excluded as valid benchmark points. We find that this behavior remains consistent across all models for $\beta < 0$ and $\alpha > 0$.
			
			In the present analysis, no optimal solutions are found for $\alpha > 0.1$. As $\alpha$ increases, the dominance of the fractional matter density tends to diminish, and the field density either becomes negative or diverges. Such behavior fails to produce an elongated matter-dominated phase at high redshifts ($z \gg 1$), rendering the model incompatible with observational data. Numerical results indicate that the model yields observationally viable solutions only when the magnitude of $\alpha$ is reduced. 
			
			While the model produces late-time acceleration for most choices of $\alpha$ and $\beta$—and in certain cases results in a future Big Crunch singularity—these scenarios often fail to generate an elongated matter-dominated phase with a positive $\Omega_{\phi}$. Thus, while the model may exhibit acceleration in the near future, it does not necessarily guarantee a consistent past evolution at high redshifts as required by observations.\footnote{We note that for lower $\xi$ and $\xi \to 0$, the model exhibits a Big Crunch singularity in the near future, as shown in Fig. \ref{fig:testing_different_xi_2}, where $h \to 0$. However, the model fails to produce the elongated matter-dominated phase vital for consistent CMB observations. In Ref. \cite{Peri15}, the authors demonstrated that a linear potential with non-minimal coupling avoids the Big Crunch; however, they did not explicitly address whether such values of $\lambda$ sustain a viable matter phase at higher redshifts.}
			
			For each parameter combination, the dynamical variables saturate at specific values, demonstrating the stability of the model during late-time epochs. In every case apart from $\xi \ne 0$, the field energy density eventually dominates over the fluid density, and the effective equation of state (EoS) approaches $-1$, signaling a stable de-Sitter attractor phase.\footnote{For this analysis, we set initial conditions at $N = 0$ rather than deep within the radiation-dominated regime ($N \sim -18$) \cite{Wolf:2025jed}. We found that the dynamics remain consistent regardless of this choice. For a detailed discussion on the impact of initial conditions, see Ref. \cite{Hussain:2025vbo}.} The corresponding evolution of the Hubble parameter $h$ also saturates near $0.8$, consistent with cosmological constant behavior.}
		
		\begin{figure*}[tbh]
			{\includegraphics[scale=0.7]{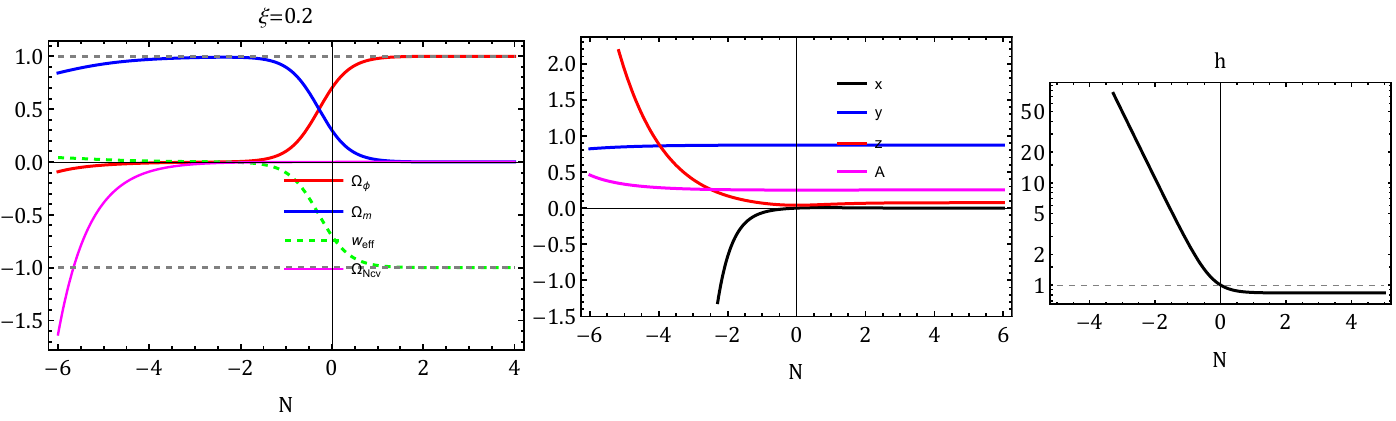}
				\includegraphics[scale=0.7]{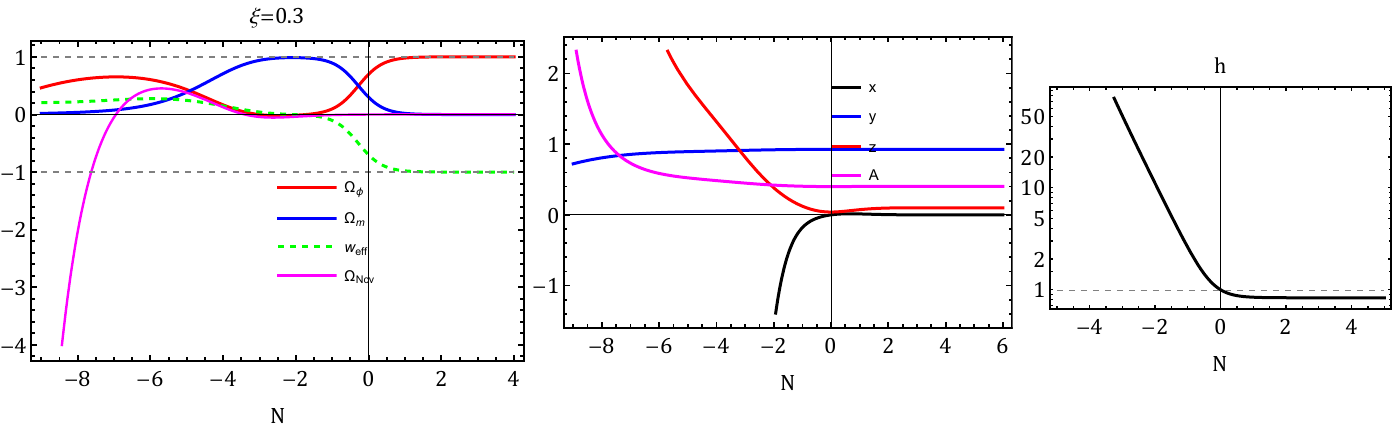}
			}		
			{	\includegraphics[scale=0.7]{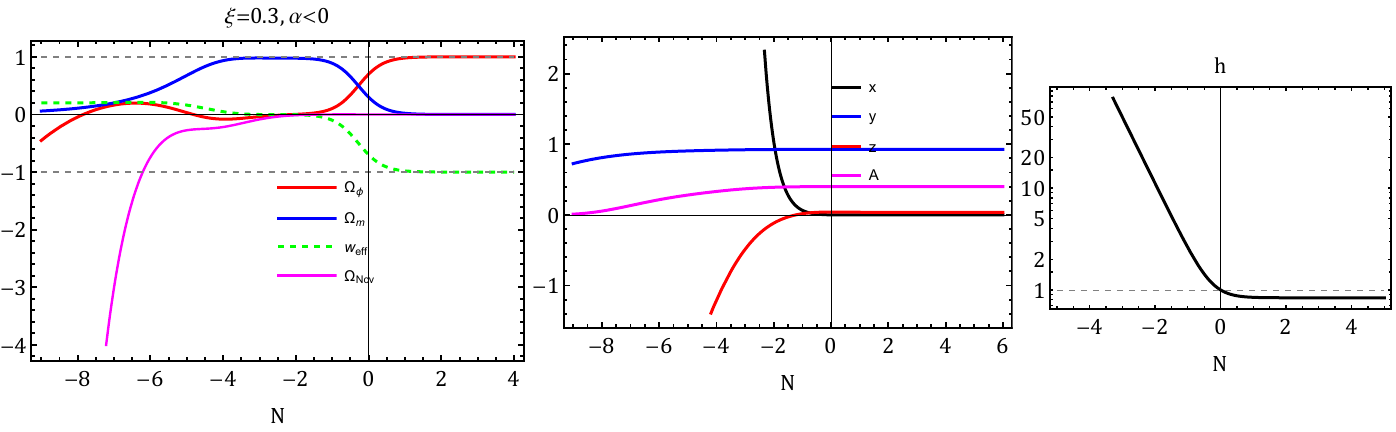}
				\includegraphics[scale=0.7]{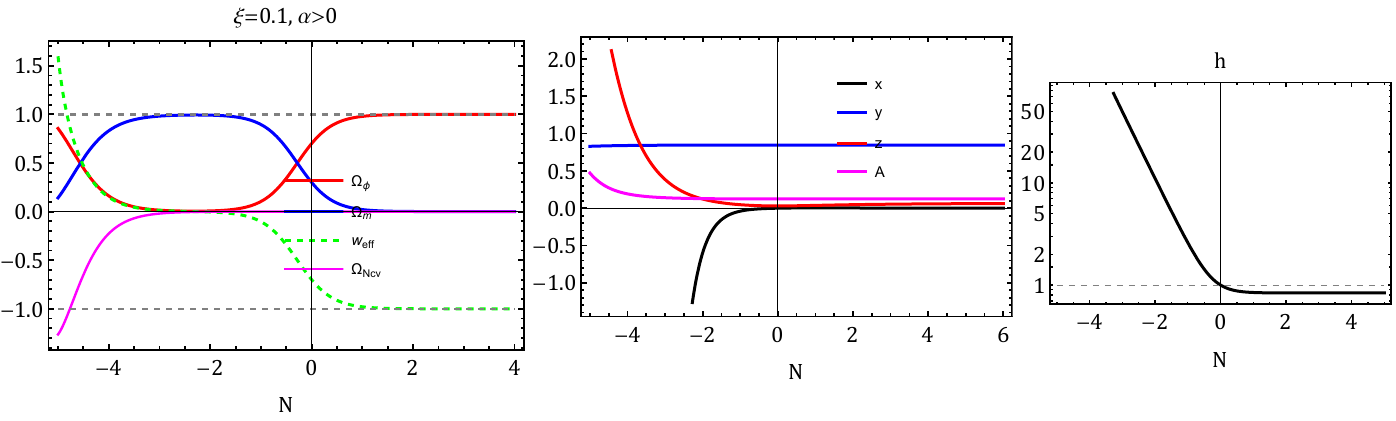} }
			
			\caption{{The evolution of the cosmological parameters and dynamical variables is shown for the parameter sets \((\xi=0.2, \alpha=0.03, \beta=0.3, x_0=10^{-4}, z_0=0.04, A_0=0.25)\), \((\xi=0.3, \alpha=0.04, \beta=0.3, x_0=10^{-4}, z_0=0.04, A_0=0.4)\), \((\xi=0.3, \alpha=-0.02, \beta=0.3, x_0=10^{-4}, z_0=0.04, A_0=0.4)\), and \((\xi=0.1, \alpha=0.01, \beta=0.3, x_0=10^{-4}, z_0=0.03, A_0=0.124)\), displayed from the top to the bottom rows, respectively. Each row corresponds to a fixed value of $\xi$, as indicated in the title of the first panel in the corresponding row. For all evolutions, we set $\Omega_{m0}=0.31$. Here, $(x_0, z_0, A_0)$ denote the present-day values of the dynamical variables at $N=0$. These parameters are chosen such that the model admits physically viable cosmological solutions throughout the redshift range considered.}}
			\label{fig:testing_different_xi_1}
		\end{figure*} 
		
		\begin{figure*}
			{\includegraphics[scale=0.7]{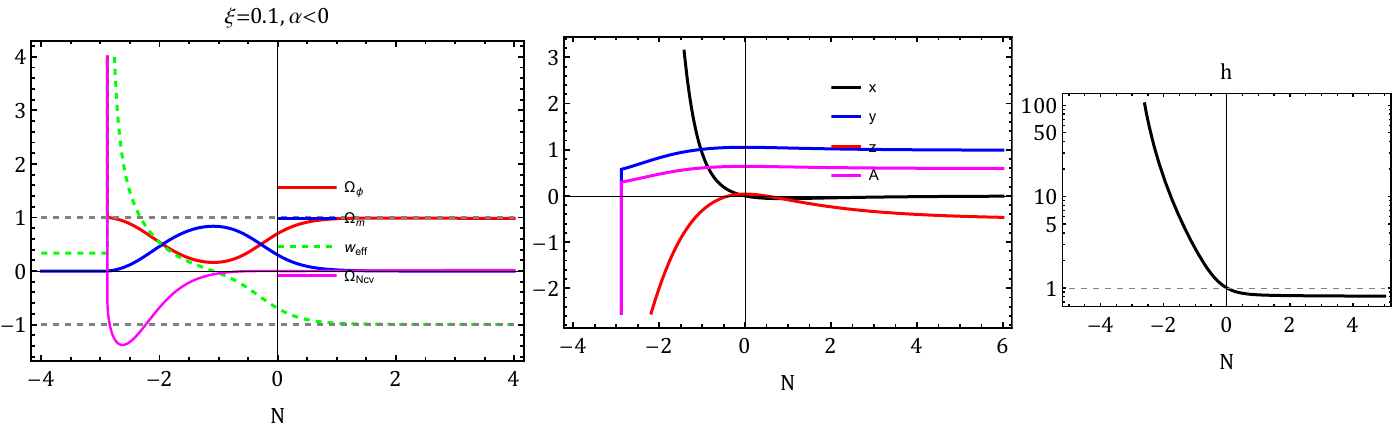}
				\includegraphics[scale=0.7]{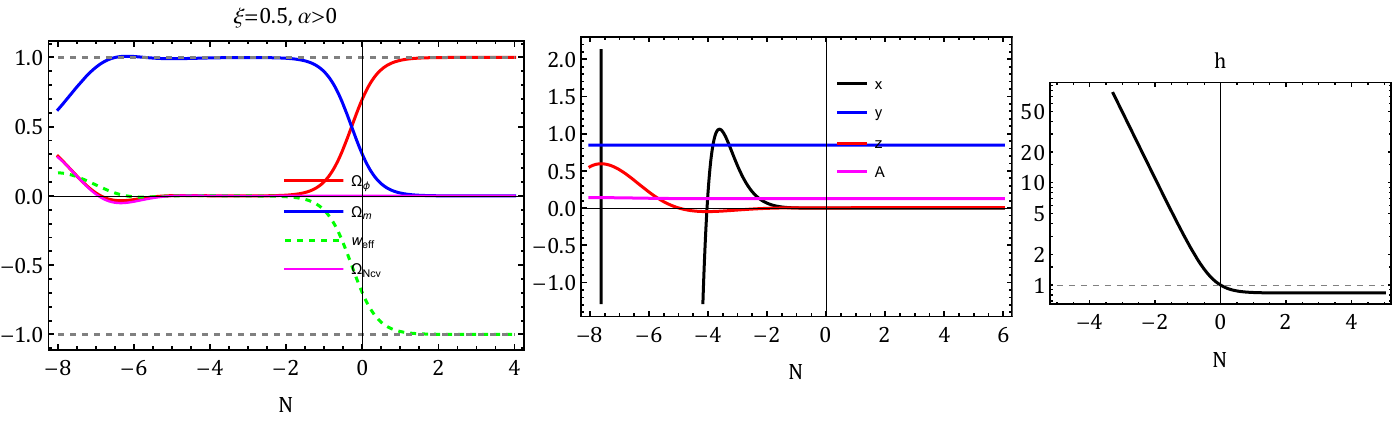}
				\includegraphics[scale=0.7]{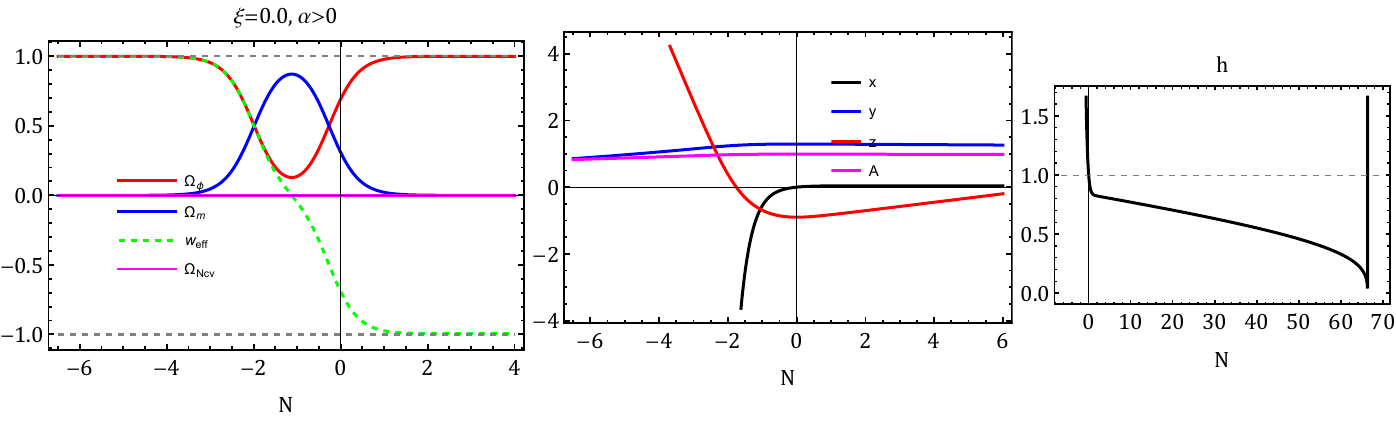} }
			\caption{The evolution of the cosmological parameters and dynamical variables for \((\xi= 0.1, \alpha = -0.24, \beta = 0.3, x_0 = 10^{-4}, z_0 = 0.03, A_0 = 0.64)\), \((\xi= 0.5, \alpha = 0.003, \beta = 0.3, x_0 = 10^{-4}, z_0 = 0.004, A_0 = 0.128)\), $(\xi= 0, \alpha = 0.069, \beta = -0.03, x_0 = 10^{-4}, z_0 = -0.9, A_0 = 0.99)$. We choose $\Omega_{m0} = 0.31$ for all the evolution. {We adopt the similar placement strategy as in Fig. \ref{fig:testing_different_xi_1}. Here, for the last panel corresponding to $\xi = 0$, we have increased the time variable $N$ up to $80$, as to specifically show that the selected parameter value yield slow contraction phase $h \to 0$, which is not observed in other cases for higher $N>0$.}  }
			\label{fig:testing_different_xi_2}
		\end{figure*}
		
		\section{Data Analysis}
		\label{sec:DA}
		\noindent In this section, we discuss the datasets used to constrain the model parameters.
		\begin{itemize}
			{
				\item \textbf{CC Data:} The Cosmic Chronometer (CC) data set contains $32$ model independent observational data points of the Hubble parameter in the redshift range \(\in [0.07,1.965]\) \cite{Vagnozzi:2020dfn,Jimenez:2001gg,Moresco:2015cya}. Out of 32 points, there are $15$ points which are highly correlated \cite{Moresco:2012jh,Moresco:2015cya,Moresco:2016mzx}. \footnote{The covariance matrix between those points can be downloaded here: \url{https://gitlab.com/mmoresco/CCcovariance}} We refer to this dataset as `CC'.
				
				\item \textbf{SN Data:} In this study, we use observational data from Type Ia Supernovae (SNe) provided by two different observations: Pantheon Plus and DES5YR \cite{Brout:2022vxf, Scolnic:2021amr,DESI:2024mwx}. The most recent update from the Pantheon+ compilation includes $1701$ cosmologically viable SNe Ia light curves from $1550$ distinct supernovae. We refer to this as the ``PP'' sample, covering the redshift range \(\in [0.001, 2.26]\). We use the likelihood provided in Ref.~\cite{Brout:2022vxf}, which incorporates the full statistical and systematic covariance matrix. From this catalog, we use the data column corresponding to the apparent magnitude of the supernovae \(m\), rather than the distance modulus \(\mu\). We apply a filter on redshift $z>0.01$, discarding nearby samples that contains higher uncertainty. We analytically marginalize absolute brightness $M_B$ by following the procedure outlined in \cite{Goliath:2001af} as implemented in Cobaya.\footnote{Likelihood estimation code: \url{https://github.com/CobayaSampler/cobaya/blob/master/cobaya/likelihoods/sn/pantheonplus.py}.}
				
				The second supernova dataset is from the Dark Energy Survey \textbf{(DES-SN5YR)}, which includes $1829$ distinct SNe \cite{DESI:2024mwx}. It consists of $194$ nearby SNe samples with redshift \(<0.1\), and $1635$ DES SNe samples. We compute the likelihood using the data column corresponding to the distance modulus \(\mu\), along with the full covariance matrix. Throughout the text, we refer to this dataset as ``DES''. However, recently the datasets have been revised with the new samples and the set called DES-Dovekie \cite{DES:2025sig}.\footnote{The updated \href{https://github.com/des-science/DES-SN5YR/tree/main/4_DISTANCES_COVMAT}{DES-Dovekie sample} and likelihood calculation can be found here -- \url{https://github.com/des-science/DES-SN5YR/blob/main/5_COSMOLOGY/Dovekie_cosmosis_likelihood.py}.} 
				
				The distance modulus is defined as:  }
			\begin{equation}
				\mu = m-M = 5\log(d_L/\text{Mpc}) + 25 \   \ .
			\end{equation}
			Here, \(m\) is the apparent magnitude of the supernova and \(d_L\) is the luminosity distance: 	
			\begin{equation}
				\label{eq:lum}
				d_L({z}) = c(1+{z}) \int_0^{{z}} \frac{dz'}{H(z')} \ ,
			\end{equation} 
			for the flat FLRW metric, where \(c\) denotes the speed of light in km/s. 
			The model parameters are constrained by minimizing a chi-square ($\chi^2$) likelihood, defined as:
			\begin{equation}
				-2 \ln (\mathcal{L}) = \chi^2 = \Delta D^{T} \mathcal{C}^{-1}_{\rm stat+sys} \Delta D\ ,
			\end{equation}
			where 
			\begin{equation}
				\Delta D = \mu_{\rm Obs} - \mu_{\rm Model} \ .
			\end{equation}
			
			\item {\textbf{Union 3 data:} The data set is also corresponding to the supernova sample which is a compilation of $2087$ SNe Ia from $24$ independent datasets \cite{Rubin:2023jdq}. The data set consists of $22$ binned samples where the catalog have been updated with full covariance matrix between these samples. We refer this data set `Uni3' throughout the text.  }
			
			\item \textbf{BAO data:} The Baryonic Acoustic Oscillation (BAO) dataset, referred to as \textbf{DESIBAO}, consists of 7 data points from the Dark Energy Spectroscopic Instrument (DESI) Release II \cite{DESI:2025zgx}, which is an updated and enhanced version of DR1 \cite{DESI:2024mwx, DESI:2019jxc, Moon:2023jgl}. We derive the constraints using the full covariance matrix provided in \cite{DESI:2025zgx}. The relevant observables include $(D_M, D_H, D_V, r_d)$, corresponding to the comoving angular distance, comoving Hubble distance, spherically averaged distance, and the sound horizon at the drag epoch, respectively. For mathematical definitions of these parameters, we refer readers to Refs.~\cite{eBOSS:2020yzd,DESI:2024mwx}. In this study, we treat $r_d$ as a free parameter, which we constrain using various combinations of datasets. The value of \(r_d\) as estimated from Planck observations using $\Lambda$CDM is \(r_d = 147.09 \pm 0.26 \) \cite{Planck:2018vyg} where as $H_0 r_d$ from DESI DR2 is approximately $(100-101) \times 100\ \mathrm{km\,s^{-1}}$, reflecting an approximate 40\% improvement over DR1. We refer this data set as `DBAO'. 
			
		\end{itemize}
		
		{In this study we treat CC+DBAO as a Baseline data set and will join this with three different types of compilation of supernovas, to compute the likelihood:

			\begin{equation}
				\begin{split}
					(i)\ 	\chi^2_{\rm tot} = \chi^2_{\rm CC} + \chi_{\rm DBAO}^2 + \chi^2_{\rm PP}  \ ,  \\ (ii) \ \chi^2_{\rm tot} = \chi^2_{\rm CC} + \chi_{\rm DBAO}^2 + \chi^2_{\rm DES}\ ,\\ (iii)\ 	\chi^2_{\rm tot} = \chi^2_{\rm CC} + \chi_{\rm DBAO}^2 + \chi^2_{\rm Uni3}  \ .
				\end{split}
			\end{equation}

			The total likelihood of the system is then computed as:
			\begin{equation}
				-2 \ln \mathcal{L}_{\rm tot} = \chi^2 _{\rm tot}\ ,
			\end{equation}
			from which we obtain the 1D and 2D posterior probability distributions of the model parameters. The posterior distribution is obtained using Bayes' theorem, which requires careful selection of prior ranges. Based on our prior discussion of the system's dynamics, we apply uniform priors as listed in Tab. \ref{tab:prior_range}. {For sampling, likelihood and log-evidence ($\log Z$) estimation, we use the publicly available dynamical nested sampler \texttt{dynesty} and the resulting samples are analyzed using \texttt{GetDist} \cite{Torrado:2020dgo,Lewis:2019xzd,Sergey:dynesty, 2020MNRAS.493.3132S, 2004AIPC..735..395S, Higson:dynamics_nested}. We compare the performance of each model against the flat $\Lambda$CDM baseline for all considered likelihoods.  }

			
			The statistical significance of the evidence difference is interpreted according to the revised Jeffreys' scale, defined by $\Delta  \log Z \equiv  \log Z_{\rm Model} -  \log Z_{\rm \Lambda CDM}$. Following the criteria in \cite{Escamilla:2024fzq,kass1995}, the results are classified as follows: $0 < |\Delta  \log Z| < 1$ is inconclusive, $1 < |\Delta  \log Z| < 2.5$ suggests weak evidence, $2.5 < |\Delta  \log Z| < 5$ indicates moderate evidence, and $|\Delta  \log Z| > 5$ represents strong evidence.

			\subsection{Results and Discussion}
			\label{sec:RES}
			
			{In this section, we confront the theoretical model with observational datasets. Following the detailed numerical investigation presented in the previous section, where the model behavior was examined for a wide range of initial conditions and parameter values, we define the prior ranges of the model parameters as listed in Tab.~\ref{tab:prior_range}. {For the observational analysis, we consider three different cases, denoted as Models I, II, and III, corresponding to fixed values $\xi=0.3$, $\xi=0.5$, and a variable coupling parameter $\xi \in [0.01,0.7]$, respectively. Although Models I and II represent specific cases of Model III and therefore introduce additional assumptions into the analysis, our numerical investigation indicates that larger values of $\xi$ tend to suppress the slow-contraction phase. At the same time, the normalized Hubble parameter $h$ exhibits a transient evolution before approaching a constant value corresponding to a stable de Sitter phase, as illustrated in Fig.~\ref{fig:xi_1_alpha_beta}. Since MCMC provides a robust statistical framework for exploring the parameter space and identifying observationally viable regions, the analysis of Models I and II can offer valuable insight into the current tensions associated with the measurement of the Hubble constant and the possible dynamical nature of dark energy. Therefore, even if current observations do not strongly favor these scenarios, Models I and II remain phenomenologically interesting and may provide testable deviations from $\Lambda$CDM that could be probed by future observations.}}

			
			Besides the model parameters $\alpha$ and $\beta$, we also vary the dynamical variables $(z_0, A_0)$ within specific ranges identified through numerical simulations, for which the model yields physically viable solutions. To better capture the sensitivity of the model parameters, we reparameterize selected quantities in logarithmic form. In particular, we define $\alpha \rightarrow 10^{\bar{\alpha}}$ and sample $\bar{\alpha} \in [-5,0.5]$, which corresponds to $\log_{10}\alpha \in [-5,0.5]$. Therefore, the physical parameter $\alpha$ is sampled logarithmically over the range $[10^{-5},10^{0.5}]$. Similarly, $z_0$ is sampled logarithmically such that $\bar{z}_0 = \log_{10}z_0 \in [-5,0.5]$, corresponding to $z_0 \in [10^{-5},10^{0.5}]$. For each case, we fix $x_0 \sim 10^{-4}$, as it was found in numerical simulation that below this range the model produces no appreciable deviation for any value of $\xi$ considered in this study. Although $A_0$ and $z_0$, corresponding to the potential and the scalar field variable $\phi$, respectively, are not directly observable quantities, in this dynamically complex system their values must lie within a specific range that simultaneously ensures the present-day field density parameter remains close to $\Omega_{\phi} \approx 0.7$ and that the past evolution yields a sufficiently prolonged matter-dominated phase. {These admissible ranges are determined from the numerical viability analysis described in Sec.~\ref{sec:EDA} (Figs.~\ref{fig:testing_different_xi_1}--\ref{fig:testing_different_xi_2}): parameter combinations yielding $\Omega_\phi < 0$ or $\Omega_m < 0$ at any redshift are excluded.\footnote{{Although the figures illustrate in the numerical simulation section are for selected benchmark parameter values, the qualitative behavior remains unchanged within their neighboring regions. The prior ranges were therefore chosen to encompass these physically relevant regions of parameter space. In addition, we impose a physical viability criterion within our Python pipeline that automatically discards parameter combinations leading to negative fractional energy densities for either the scalar field or the matter component. Owing to this implementation, we adopt relatively broad flat priors for $A_0$ and $\bar{z}_0$.}}} Together with the model parameters, these dynamical variables therefore ensure that the overall dynamics remain physically viable across all redshifts. Consequently, while $(z_0, A_0)$ are sampled through the MCMC analysis, their admissible ranges are determined entirely by physical viability conditions established from numerical simulations, rather than by data-fitting considerations; they therefore do not act as phenomenological free parameters introduced to artificially improve the fit.
			
			Owing to the highly non-linear dynamics of the model, the field fractional density $\Omega_{\phi}$ may become negative at high redshifts. Consequently, we implemented a rigorous constraint within the MCMC Python pipeline: any parameter combination yielding a negative value for $\Omega_{\phi}$ or $\Omega_{m}$ at any point in the redshift evolution results in a discarded likelihood. Such points are classified as physically non-viable. Although the observational datasets extend only to $z < 3.0$, we numerically integrate the coupled differential equations up to $N = -6.0$ to ensure that the cosmological parameters remain physically consistent throughout the early-time evolution.
			
			For this analysis, the initial condition for the kinetic component of the field is fixed at $x(0) = 10^{-4}$, as values below this threshold do not induce significant variations in the cosmological dynamics. The marginalized 1D and 2D posterior distributions of the model parameters for the considered datasets are illustrated in the corner plot in Fig. \ref{fig:corner_plot}, while the corresponding mean values at the $68\%$ confidence level (CL) are summarized in Tab. \ref{tab:mean_value_para}.    }
		
		\begin{table}
			\centering
			
			\resizebox{\columnwidth}{!}{	\begin{tabular}{l cccccc c c }
					\hline
					\multicolumn{9}{c}{\bf\boldmath Prior Range for Models I--$\xi = 0.3$, II--$\xi=0.5$}\\
					\hline
					\hline
					{\bf } & $H_0$ & $\Omega_{m}$ & $r_d$ & \(\bar{\alpha}\) & $\beta$ & ${A_0}$ & $\bar{z}_0$ & $x_0$ \\
					\hline
					& $[30,100]$ & $[0.0, 0.7]$ & $[100, 300]$ & $[-5,0.5]$ & $[0.0,0.8]$ & $[0.001, 0.75]$ & $[-5, 0.5]$ & $10^{-4}$ \\
					\hline 
					\hline
					\multicolumn{9}{c}{\bf Model III}\\
					\hline
					\multicolumn{9}{c}{\boldmath $\xi: \ [0.01, 0.7] $}\\
					\hline
					\hline
					
			\end{tabular}}
			\caption{The uniform prior ranges for Models I, II, and III. Here, we re-scaled the parameters $\alpha, z_0$ to $10^{\bar{\alpha}}$ and $10^{\bar{z}_0}$ respectively and constraining $\bar{\alpha}$ and $\bar{z}_0$ to capture the sensitivity of the parameter. The prior ranges for $A_0$ and $z_0$ are informed by the physical viability analysis of Sec.~\ref{sec:EDA} rather than purely by data-fitting.}
			\label{tab:prior_range}
		\end{table}

		\begin{table*}
			\centering
			
			\resizebox{\linewidth}{!}{\begin{tabular}{l cccccc c c r}
					\hline
					\multicolumn{10}{c}{\bf Datasets: CC+DBAO+PP}\\
					\hline
					\hline
					{\bf Models} & $H_0$ & $\Omega_{m}$ & $r_d$ & \(\bar{\alpha}\) & $\beta$ & ${A_0}$ & $\bar{z}_0$ & $\xi$ & $\log Z \pm \sigma$\\
					\hline
					{\bf Model I:} & $70.0^{+1.9}_{-2.2}$ & $0.2968^{+0.0082}_{-0.0071}$ & $143.9\pm 4.3$ &$-3.7^{+1.8}_{-1.6}$ & $ 0.447^{+0.093}_{-0.16}$ & $0.549^{+0.065}_{-0.093}$ & $-1.051^{+0.049}_{-0.063}$&  & $-882.49 \pm 0.13$ \\

					{\bf Model II:} & $70.2 \pm 2.3	$ & $0.2560^{+0.0073}_{-0.0062}$ & $145.5^{+4.2}_{-5.2}$ & $-3.98^{+1.8}_{-0.88}$ & $0.554^{0.043}_{-0.039}$ & $0.670	^{+0.027}_{-0.014}$ & $-1.027^{+0.013}_{-0.018}$ & & $-907.85 \pm 0.19$ \\
					
					{\bf Model III:} & $69.0 \pm 2.1	$ & $0.3137\pm 0.0081$ & $145.3^{+4.1}_{-4.8}$ & $-4.57^{+0.64}_{-1.1}$ & $0.37^{0.16}_{-0.25}$ & $0.072	^{+0.018}_{-0.030}$ & $-3.66^{+0.56}_{-1.2}$ & $0.131^{+0.035}_{-0.0082}$  & $-875.74 \pm 0.10$  \\
					
					{\boldmath\bf $\Lambda$CDM:} &$69.2 \pm 2.4$ & $0.3124 \pm 0.0081$& $145.2\pm 5.0$&&&&&& $-869.96 \pm 0.15$\\
					
					\hline
					\multicolumn{10}{c}{\bf Datasets: CC+DBAO+DES}\\
					\hline
					\hline
					{\bf Model I:} & $70.0^{+2.2}_{-2.4}$ & $0.2942 \pm 0.0079$ & $144.4\pm 4.7$ &$-4.0^{+1.5}_{-1.7}$ & $ 0.407 ^{+0.059}_{-0.11}$ & $0.567\pm 0.057$ & $-1.067^{+0.044}_{-0.066}$ & & $-856.83 \pm 0.14$ \\
					
					{\bf Model II:} & $70.2 \pm 2.2$ & $0.2532\pm 0.0067$ & $145.8^{+4.2}_{-4.8}$ & $-4.2^{+1.0}_{-1.6}$ & $0.552^{+0.047}_{-0.041}$ & $0.670^{+0.029}_{-0.013}$ & $-1.028^{+0.014}_{-0.021}$ & & $-879.13 \pm 0.19$\\
					
					{\bf Model III:} & $69.8 ^{+2.3}_{-2.0}	$ & $0.3095\pm 0.0078$ & $144.2^{+3.8}_{-5.0}$ & $-4.1^{+1.6}_{-1.8}$ & $0.23^{+0.16}_{-0.36}$ & $0.059	^{+0.010}_{-0.057}$ & $-3.41^{+0.93}_{-1.0}$ & $0.098^{+0.069}_{-0.077}$  & $-851.63 \pm 0.11$  \\
					
					{\boldmath\bf $\Lambda$CDM:} &$69.3 \pm 2.4$ & {$0.3083 \pm 0.0078$}& $145.3\pm 4.9$&&&&& & $-845.29 \pm 0.15$\\
					
					\hline
					\multicolumn{10}{c}{\bf Datasets: CC+DBAO+Union3}\\
					\hline
					\hline
					{\bf Model I:} &  $69.9\pm 2.3$ & $0.2912 \pm 0.0087$ & $144.9\pm 4.7$ &$-3.62^{+1.7}_{-0.79}$ & $ 0.434 ^{+0.059}_{-0.084}$ & $0.545^{+0.046}_{-0.057}$ & $-1.067^{+0.050}_{-0.063}$ & & $-49.64 \pm 0.13$ \\
					
					{\bf Model II:} &$71.2\pm 2.3$  & $0.2476^{+0.0076}_{-0.0066}$&  $144.5^{4.1}_{-5.3}$ & $-4.0 \pm 1.1$ & $0.550^{+0.034}_{-0.040}$ & $0.668^{+0.028}_{-0.014}$ & $-1.031^{+0.015}_{-0.023}$& & $-66.69 \pm 0.17$ \\
					
					{\bf Model III:} & $69.5 \pm 2.4	$ & $0.3095\pm 0.0090$ & $144.8^{+4.5}_{-5.6}$ & $-4.18^{+1.1}_{-0.91}$ & $0.19^{0.16}_{-0.28}$ & $0.1163	^{+0.0098}_{-0.11}$ & $-2.81\pm 0.95$ & $0.122^{+0.046}_{-0.054}$  & $-44.82 \pm 0.11$  \\
					
					{\boldmath\bf $\Lambda$CDM:} &$69.6 \pm 2.4$ & $0.3065 \pm 0.0084$& $145.1\pm 4.9$&&&&& & $-37.95 \pm 0.16$ \\
					
					\hline
					\hline
			\end{tabular}}
			
			\caption{{We report the parameter constraints at the $68\%$ confidence level obtained from the MCMC analysis. The $\log Z \pm \sigma$ column gives the Bayesian log-evidence and its associated uncertainty estimated using \texttt{dynesty}. }}
			\label{tab:mean_value_para}
		\end{table*}

		\begin{figure}[t]
			
			\includegraphics[scale=0.54]{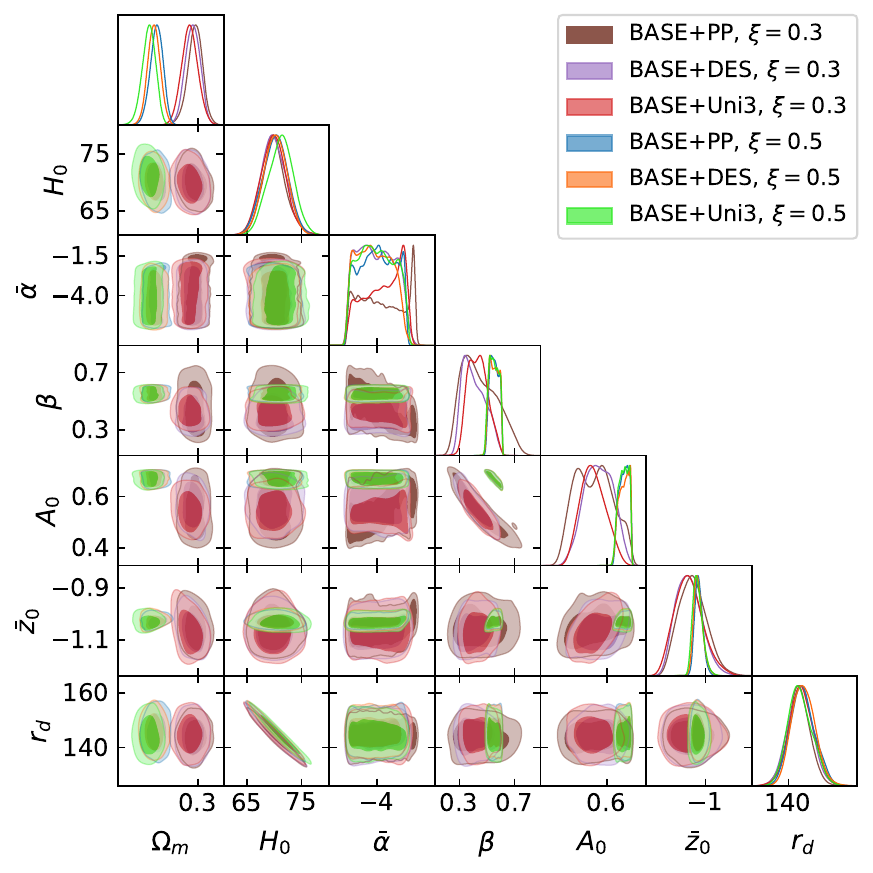}
			\includegraphics[scale=0.54]{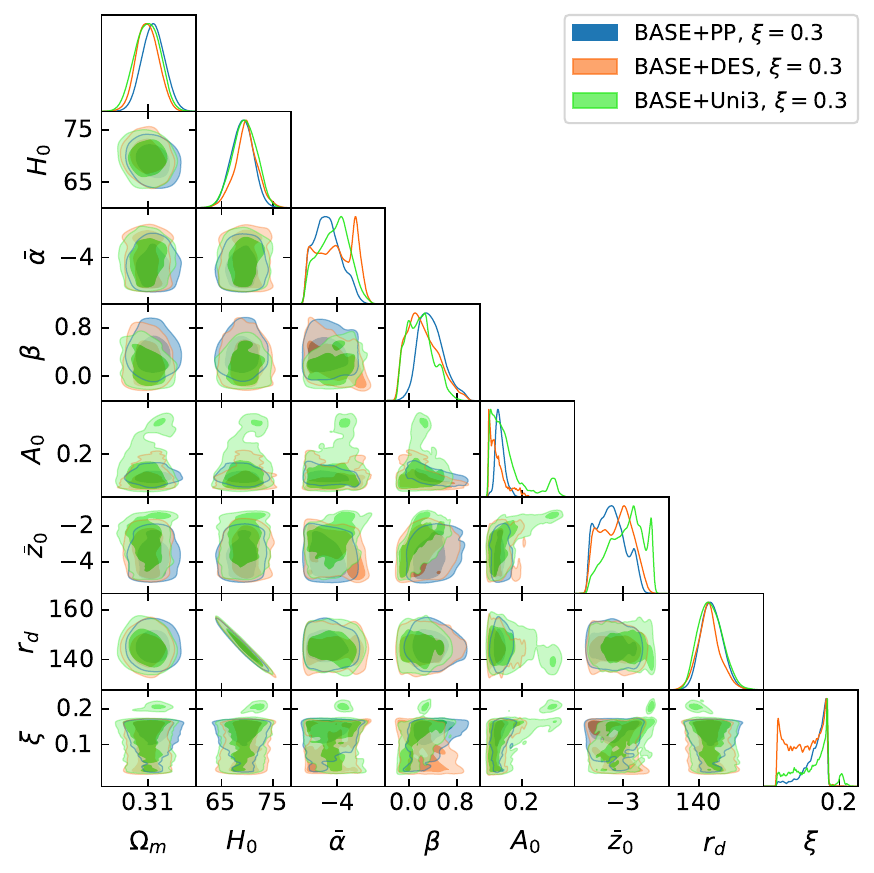}
			
			\caption{The corner plots for decaying dark-energy Models I and II are shown in the upper panels, while those for Model III are displayed in the lower panels. }
			\label{fig:corner_plot}
		\end{figure} 
		
		We find that Model I yields a little higher Hubble constant value, $H_0 \sim 70.0$ km/s/Mpc, across all datasets compared to the baseline $\Lambda$CDM model, accompanied by a lower matter density, $\Omega_{m} \sim 0.295$. Furthermore, Model I has a significantly lower Bayesian evidence than the baseline model and is therefore disfavored by the data. To constrain the remaining parameters, we re-scale variables such as $\alpha \to 10^{\bar{\alpha}}$ and $z_0 \to 10^{\bar{z}_0}$, and constraint the re-scaled parameters $\bar{\alpha}$ and $\bar{z}_0$ to better capture parameter sensitivity. We find that $\bar{\alpha } = \log_{10}\alpha \approx -3.7$ across all datasets, while $\beta \approx 0.42$. As shown in the posterior distribution in Fig. \ref{fig:corner_plot}, the $H_0$ contour spans $65$--$75$ km/s/Mpc. However, we caution that Model~I fixes the coupling parameter to $\xi=0.3$, which, for a wide range of initial conditions and model parameters, exhibits an evolution distinct from the baseline model as seen from Figs. \ref{fig:evo_total} and \ref{fig:Om_diagnostic}. Although this model yields a higher value of $H_0$, however, exhibits large magnitude of Bayesian evidence relative to the baseline model for all the considered datasets. Therefore, while the model may appear to alleviate the Hubble tension, its overall statistical performance does not provide strong support for such an interpretation. 
		
		Similar results are observed for Model II, where the model yields $H_0 \sim 71.2$ km/s/Mpc with Union3 data and remains near $70$ km/s/Mpc with other datasets. This model consistently produces a lower matter density, $\Omega_{m} \sim 0.255$, and $r_d \sim 145$ Mpc. Notably, the sound horizon remains comparable to the baseline model despite the higher $H_0$. The Model yields a higher logarithmic evidence compared to the other non-baseline models. {The large difference in the $\log Z$ values between Models I and II across all datasets further indicates that increasing $\xi$ leads to highly non-linear modifications of the cosmological evolution, producing dynamics that deviate significantly from those of $\Lambda$CDM. This effect is particularly evident in the evolution of the matter density parameter, as shown in Figs.~\ref{fig:evo_total} and \ref{fig:Om_diagnostic}. Consequently, larger values of $\xi$ are increasingly disfavored by the current combination of observational datasets. Therefore, the model-dependent increase in the inferred value of $H_0$ should not be interpreted as a genuine resolution of the Hubble tension.}

		{In Model III, where $\xi$ is allowed to vary within the range $[0.01,0.7]$, all datasets prefer relatively small values, $\xi \sim 0.12$, and yield substantially lower logarithmic evidence than their fixed-$\xi$ counterparts. As discussed previously, smaller values of $\xi$ are often associated with a future slow-contraction phase, which generally fails to produce a sufficiently prolonged matter-dominated era. However, Fig.~\ref{fig:xi_1_alpha_beta} also demonstrates that certain combinations of $({\alpha},\beta)$ can generate an extended matter-dominated phase while exhibiting a cosmological evolution closely resembling the baseline model. Indeed, the evolutions shown in Figs.~\ref{fig:evo_total} and \ref{fig:Om_diagnostic}, corresponding to the parameters constrained at the $68\%$ confidence level, display a prolonged matter-dominated epoch and closely mimic the $\Lambda$CDM behavior at higher redshifts. Furthermore, the model yields a Bayesian evidence relative to $\Lambda$CDM of $\Delta \log Z < -5$, indicating strong disfavor for the current model. However, since the evolution of the cosmological parameters closely aligns with that of the baseline model, further investigation at the perturbation level along with more comprehensive datasets, including full cosmic microwave background (CMB) observations is required to assess whether the model can serve as a robust alternative to the standard cosmological scenario.
			
			{We illustrate the evolution of cosmological parameters with respect to redshift $z$ in Fig. \ref{fig:evo_total} for the parameter constraint values at 68\% CL from Tab. \ref{tab:mean_value_para}}. The matter density tends to dilute at higher redshifts for Models I and II, with the magnitude of dilution being more pronounced for $\xi = 0.5$. In contrast, for Model III using DES and Union3 data, $\Omega_{m}$ remains close to unity at high redshifts, mimicking the baseline model. We also plot the non-minimal curvature density, $\Omega_{\rm NCV}$, which exhibits significant variation for $\xi=0.5$ while remaining near zero for the other models, indicating a strong deviation from the cosmological constant. The effective EoS is $\sim -0.5$ at the current epoch $(N=0)$, vanishes for a certain interval at intermediate redshift, and eventually saturates at $1/3$ at higher redshift for Models I and II. The observed deviations in matter density likely arise because Planck observations were not included; incorporating Planck data would likely suppress these deviations, as those observations strictly require a full matter-dominated phase. For Model III, the effective EoS which has been plotted with respect to $N$, remains zero at higher redshifts $N<0$, representing a pure matter-dominated phase, while saturated to $-1$ upon extrapolating to the late-time future epoch, demonstrating the stable de-Sitter behavior of the model. 
			
			Additionally, we plot the equation of state of the field, $w_{\phi}$, which reaches $-1$ at low redshifts and increases toward $1/3$ at higher redshifts. This transition toward $1/3$ is more abrupt for $\xi \sim 0.5$, suggesting that the field itself can behave as a radiation fluid during the radiation-dominated era. 
			
			At low redshifts, we provide a visual comparison of the Hubble evolution $H(z)$ against 32 Cosmic Chronometer (CC) data points. In the late-time regime, all models exhibit similar behavior and are nearly indistinguishable from observations alone. Finally, we show the variation of the dynamical variables $(x, \tilde{y}^2, z)$ with respect to $N = \ln a$. The variables do not diverge in the past and saturate at specific values in the future. The potential parameter difference, $\tilde{y}^2 \equiv y^2 - A^2$, remains positive throughout the evolution. Since $\tilde{y}^2$ effectively encodes the net contribution of the scalar field potential, this behavior indicates that the potential does not evolve toward negative values. As a result, the model does not enter the regime associated with a negative potential, which is typically responsible for future recollapse or Big Crunch--type singularities. Therefore, within the observationally allowed parameter space, the late-time evolution remains regular and approaches a stable accelerated phase.
			
		}
		
		\begin{figure}[t]
			\centering
			\resizebox{\columnwidth}{!}{\includegraphics[scale=0.5]{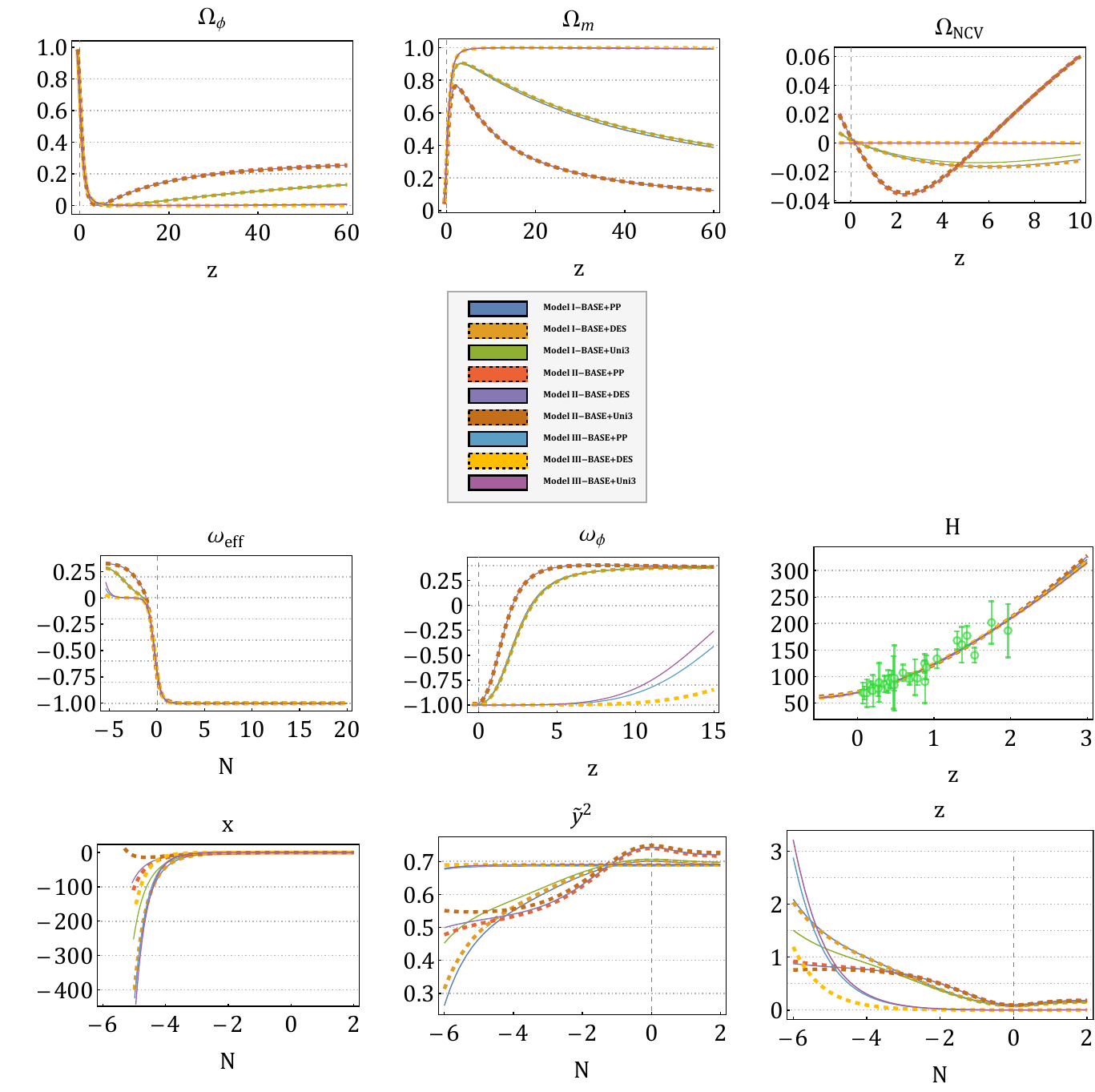}}	
			\caption{Evolution of the cosmological parameters and dynamical variables corresponding to the parameter constraint value at 68\% CL obtained from the MCMC analysis tabulated in Tab. \ref{tab:mean_value_para}. The dynamical variables $(x, \tilde{y}^2, z)$ and effective EoS $w_{\rm eff}$ are plotted as functions of the redefined time variable $N$, while the cosmological parameters are shown as functions of redshift. Additionally, the Hubble parameter evolution is plotted up to redshift $3$ against the CC dataset for visual comparison, where the green points represent the uncertainties in CC data.}
			\label{fig:evo_total}
		\end{figure}

		\begin{figure}[t]
			\resizebox{\columnwidth}{!}{\includegraphics[scale=0.6]{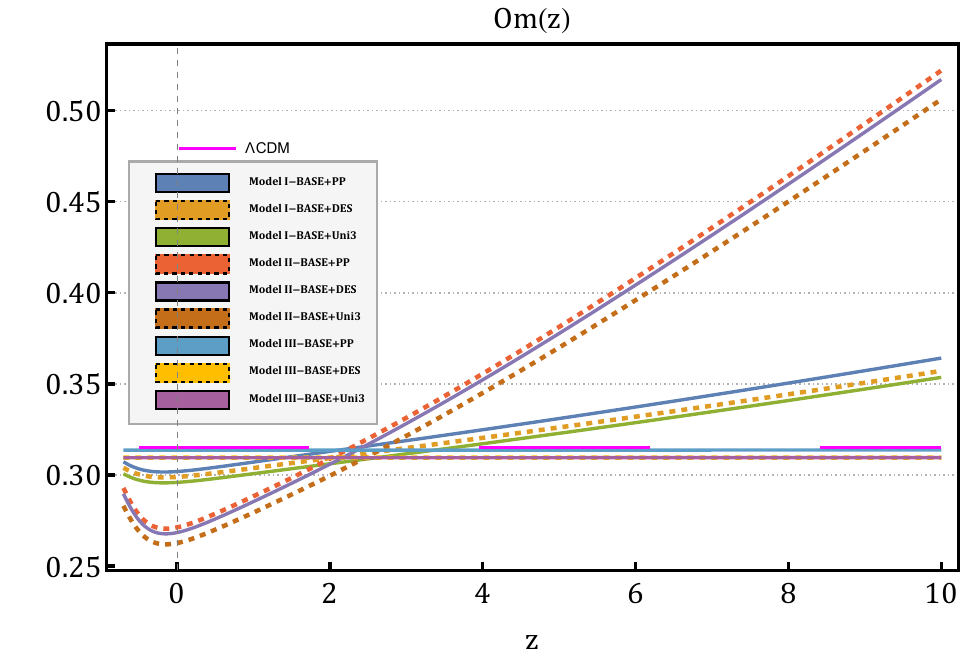}}
			\caption{The $\mathcal{O}m(z)$ diagnosis of the current model for distinct data sets is plotted against the redshift \(z\) and compared with $\Lambda$CDM.}
			\label{fig:Om_diagnostic}
		\end{figure}
		We further evaluate an important diagnostic, known as the $Om(z)$, which is particularly effective in distinguishing dynamical dark energy models from the cosmological constant in the low-redshift regime. The $Om(z)$ function is defined as \cite{Sahni:2008xx,Shahalam:2015lra,Zunckel:2008ti}.
		\begin{equation}
			Om(z) = \frac{h^2(z) - 1}{(1+z)^3 - 1}, \quad h(z)\equiv  H(z)/H_0\ .
		\end{equation}
		For a spatially flat $\Lambda$CDM model, $Om(z)$ reduces to a constant equal to $\Omega_{0m}$. In contrast, deviations from constancy in $Om(z)$ signify the presence of dynamical dark energy.
		{We display the $Om(z)$ diagnostic for all considered models in Fig. \ref{fig:Om_diagnostic}, where Model II exhibits the most significant deviation. In the low-redshift regime ($z < 2$), this model yields considerably lower value than the baseline $\Lambda$CDM model; however, the magnitude of this deviation increases at higher redshifts. In contrast, the remaining models show significantly lower deviations and remain close to the $\Lambda$CDM behavior, highlighting the dynamical nature of the proposed framework.
			
		}
		
		Regarding the phantom divide crossing in DESI DR2 BAO data, we emphasize that our model does not exhibit phantom behavior. Unlike some parameterized dark energy models that cross into the phantom regime, the present scalar field framework—owing to its positive kinetic term—remains free from ghost instabilities and maintains a consistent accelerating solution. This is an important phenomenological distinction, since it shows that the model is compatible with DESI DR2 data without invoking phantomness, while still allowing mild deviations from $\Lambda$CDM that can be probed in future surveys. Thus, even though the model does not resolve all current cosmological tensions, it enriches the landscape of viable alternatives to $\Lambda$CDM, highlights the role of non-minimal coupling in shaping cosmic evolution, and provides a physically motivated, phantom-free framework.
		
		{In non-minimally coupled (NMC) quintessence models, the coupling term $F(\phi)R$ modifies the effective gravitational interaction and may, in principle, induce fifth-force effects as well as variations in the effective gravitational constant. The precise bound on $\xi$ is model dependent; therefore, no universal fifth-force constraint on $\xi$ exists. Nevertheless, the conformal value $\xi=1/6\simeq0.167$ remains theoretically well motivated \cite{xi1,xi2,Faraoni:2000wk}, while cosmological observations generally favor positive values of $\xi$ of order $10^{-1}$ \cite{Hrycyna:2015vvs,Adam:2026ajg,Zhang:2023nil,25a3}. In the present work, the observational analysis strongly disfavors large coupling strengths and constrains the parameter to $\xi \sim 0.12$, a value that is consistent with existing fifth-force considerations. }
		
		We also note that scalar--tensor theories with the coupling form $F(\phi)=1-\kappa^2\xi\phi^2$ have been extensively studied in the literature, and compatibility with local gravity constraints generally requires either small effective couplings today or sufficiently slow field evolution. In our model, the scalar field evolves slowly at late times and asymptotically approaches a stable de-Sitter attractor, which suppresses rapid variations of the effective gravitational coupling. Moreover, the parameter region favored by cosmological observations in our analysis corresponds to relatively small nonminimal coupling strengths. Therefore, the observationally allowed region obtained in this work is expected to remain compatible with current solar-system bounds on deviations from general relativity.
	}\\
	
	\section{Phase Space Analysis: Stationary points and their stability}
	\label{sec:phase}
\begin{figure}[tbh]
		\resizebox{\columnwidth}{!}{\includegraphics{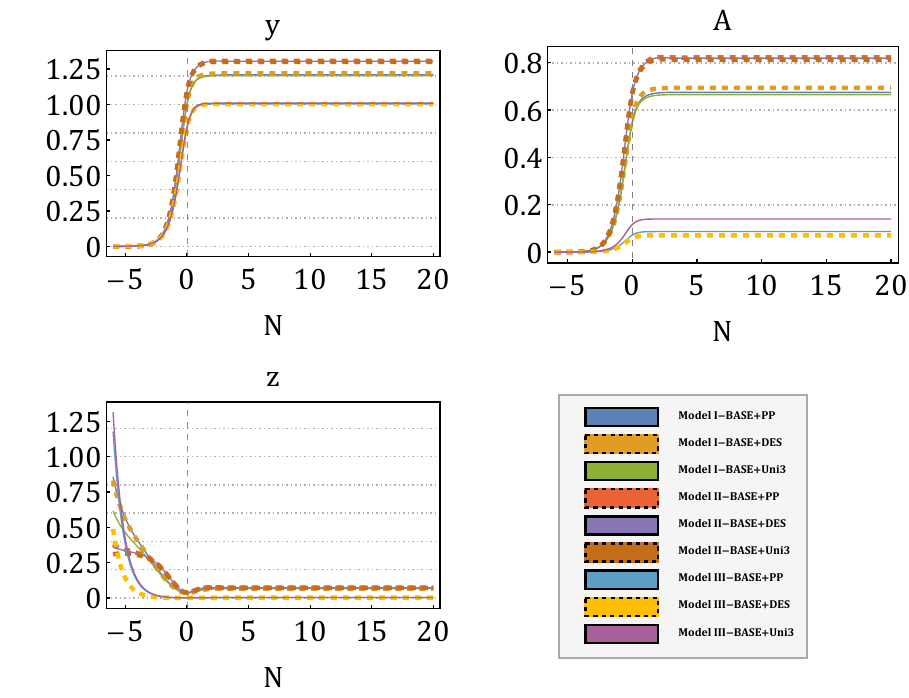}
		}
		\caption{Evolution of the potential-related variables $y$, $A$ and $z$, defined in Eq.~\eqref{eq:var}, for the parameter constraint listed in Tab. \ref{tab:mean_value_para}. For all models, the variables asymptotically evolve toward values satisfying $y>A$, thereby avoiding the future slow-contraction phase. The dynamical variables converge to the stable point $A_{5}$. Consequently, the cosmological evolution approaches a stable de Sitter phase with $w_{\rm eff}\rightarrow -1$, consistent with the behavior shown in Fig.~\ref{fig:evo_total}.}
		\label{fig:dyn_stab_evo}
	\end{figure}

\begin{figure}
		\resizebox{\columnwidth}{!}{\includegraphics{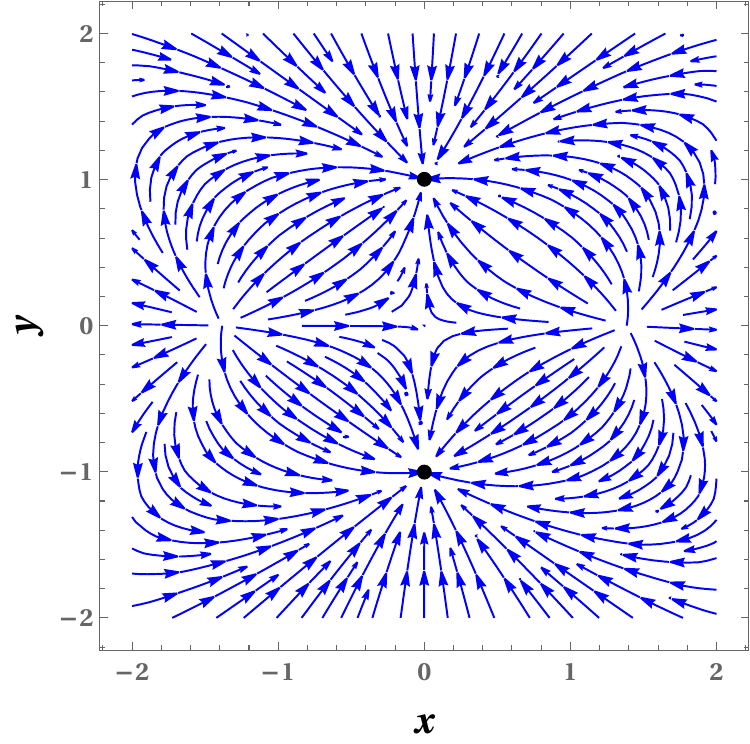}}	
		\caption{{The figure illustrates the reduced phase space for the stable fixed point $A_5$, with the planes fixed at $A=0.1$ and $z=0.001$. The analysis is performed using the observationally favored parameter values $\alpha \sim 10^{-4}$, $\beta \simeq 0.26$, $\xi \simeq 0.12$, and $w_m=0$.  The black dots represent stable attractor points.}}
		\label{fig:port}
	\end{figure}

		Phase space analysis serves as a mathematical tool for examining the dynamics of systems by assessing the stability of critical points via eigenvalues and illustrating the trajectories within the phase space \cite{Shahalam:2026abt,Myrzakulov:2025ogj,Shahalam:2026obq}. It aids in comprehending how cosmological models evolve and remain stable. The fixed point can be classified as stable or unstable, depending on whether the real part of the eigenvalues are positive or negative. The eigenvalues at a certain fixed point can occasionally show both positive and negative signs. We call this location a saddle point. {In this section, we assess the background-level stability of the model using the parameter constraints obtained in Tab.~\ref{tab:mean_value_para}. Here, we primarily focus on Model III $(\xi \sim 0.12)$, as it is the most observationally close scenario relative to the baseline model among all the variants considered. To facilitate the dynamical-system analysis, we introduce the following dimensionless variables: } 
	\begin{equation}
		x=\frac{\kappa \dot{\phi}}{\sqrt{6}H}, \ y=\frac{\kappa \sqrt{V_0 e^{-\kappa \alpha \phi}}}{\sqrt{3}H}, \ A=\frac{\kappa \sqrt{V_1 e^{\kappa \beta \phi}}}{\sqrt{3}H}, \ z=\frac{\kappa \phi}{\sqrt{6}},
		\label{eq:var}
	\end{equation}
	{Note that these variables have a structure similar to those defined in Eq.~\eqref{dimensionless_variable_data}, with the key difference that the field variables are now normalized with respect to the time-dependent Hubble parameter $H$.} The equations of motion (\ref{eq:Friedphi})$-$(\ref{eq:KGphi}) are reformulated in new variables as a system of differential equations of first order.
	\begin{eqnarray}
		\frac{dx}{dN} &=& x \Big{(}\frac{\ddot{\phi}}{H \dot{\phi}} - \frac{\dot{H}}{H^2}\Big{)}, \nonumber \\
		\frac{dy}{dN} &=& -y \Big{(} \frac{\sqrt{6}x \alpha}{2} + \frac{\dot{H}}{H^2}\Big{)}, \nonumber \\
		\frac{dA}{dN} &=& A \Big{(} \frac{\sqrt{6}x \beta}{2} - \frac{\dot{H}}{H^2}\Big{)}, \nonumber \\
		\frac{dz}{dN} &=& x,
		\label{eq:auto}
	\end{eqnarray}
	where
	\begin{widetext}
		\begin{eqnarray}
			N&=& \ln a,\\
			\frac{\dot{H}}{H^2} &=& -\frac{4+4 A^2+2 x^2-4 y^2-12 x^2 \xi -24 z^2 \xi +6 \sqrt{6} y^2 z \alpha  \xi +6 \sqrt{6} A^2 z \beta  \xi +144 z^2 \xi ^2-\Omega_m+3 w_m \Omega_m}{2 \left(1-6 z^2 \xi +36 z^2 \xi ^2\right)}, \label{eq:Hd}\\
			\frac{\ddot{\phi}}{H \dot{\phi}} &=& -\frac{1}{2 x \left(1-6 z^2 \xi +36 z^2 \xi ^2\right)} \bigg(6 x+\sqrt{6} y^2 \alpha +\sqrt{6} A^2 \beta -24 A^2 z \xi -12 x^2 z \xi +24 y^2 z \xi -36 x z^2 \xi  \nonumber \\
			&& -6 \sqrt{6} y^2 z^2 \alpha  \xi  -6 \sqrt{6} A^2 z^2 \beta  \xi +72 x^2 z \xi ^2+216 x z^2 \xi ^2+6 z \xi  \Omega_m-18 w_m z \xi  \Omega_m \bigg),
		\end{eqnarray}
	\end{widetext}
	\begin{equation}
		\Omega_m = 1-(x^2 + y^2 - A^2 + 12 \xi x z + 6 \xi z^2). \label{frdn_eq}
	\end{equation}
	We employ the autonomous system (\ref{eq:auto}) to determine the stationary points by equating the left-hand side of these equations to zero. Consequently, we identify the following stationary points labeled as $A_{i}$:
	\begin{eqnarray}
		\label{eq:point1}
		A_1: \quad 	x=0, \qquad y=0, \qquad A=0, \qquad z=0.
	\end{eqnarray}
	The corresponding eigenvalues are given by
	\begin{multline}
		\label{eq:A1}
		{\mu}_1 =3(1+w_m)/2, \   
		{\mu}_2 =3(1+w_m)/2, \\ 
		{\mu}_3 = - (3-3w_m+\sqrt{9-18w_m+9w_m^2-48 \xi+144 \xi w_m})/4, \\
		{\mu}_4 = - (3-3w_m-\sqrt{9-18w_m+9w_m^2-48 \xi+144 \xi w_m})/4\ .
	\end{multline}	
	Two eigenvalues are negative and the other two are positive. Therefore, this is a saddle point.
	\begin{eqnarray}
		\label{eq:point2}
		A_2: \quad 	x=0, \qquad y=0, \qquad A=0, \qquad z= \pm \frac{1}{\sqrt{6 \xi}}\ .
	\end{eqnarray}
	In this case, the eigenvalues are
	\begin{multline}
		\label{eq:A2}
		{\mu}_1 =2,  
		{\mu}_2 =2, 
		{\mu}_3 = - 1,
		{\mu}_4 = 1-3 w_m\ .
	\end{multline}
	This point is also saddle as one of the eigenvalues is negative and rests are positive.
			\begin{multline}
			\label{eq:point3}
			A_3: \quad 	x=0, \ y=\pm \frac{2\sqrt{-2\xi+\sqrt{\xi(\alpha^2+4\xi)}}}{\alpha}, \ A=0, \\ z=  \frac{2\sqrt{6}\xi \pm \sqrt{6\alpha^2 \xi+24 \xi^2}}{6 \alpha \xi}.
		\end{multline}
		{The corresponding eigenvalues for the observationally preferred parameters $\alpha \sim 10^{-4}$, $\beta \simeq 0.26$, $\xi \simeq 0.12$, and $w_m=0$ are given by,
			\begin{multline}
				\label{eq:A3}
				{\mu}_1 = -3.0, 
				{\mu}_2 = -2.4,  
				{\mu}_3 = -0.6, 
				{\mu}_4 =  -2.7 \times 10^{-9}.
			\end{multline}
			The fourth eigenvalues at the fixed points $A_3$ is extremely close to zero ($\mu_4 \sim 10^{-9}$) and may indeed be numerical artifacts arising from machine precision rather than genuinely negative eigenvalues. Since one eigenvalue of the fixed point vanishes, the point $A_3$ is formally a non-hyperbolic fixed point.
			\begin{eqnarray}
				\label{eq:point4}
				A_4: \quad x=0,  y=0,  A=\frac{2\sqrt{2\xi+\sqrt{\xi(\beta^2+4\xi)}}}{\beta}, \nonumber\\ z=  \frac{-2\sqrt{6}\xi + \sqrt{6\beta^2 \xi+24 \xi^2}}{6 \beta \xi}.
			\end{eqnarray}
			The corresponding eigenvalues for data favored parameter values $\alpha \sim 10^{-4}$, $\beta \simeq 0.26$, $\xi \simeq 0.12$, and $w_m=0$ are given by,
			\begin{equation}
				\label{eq:A4}
				{\mu}_1 = 139.2,  {\mu}_2 = 47.4, {\mu}_3 = 45.2, {\mu}_4 =  -0.76.
			\end{equation}
			This point is classified as a saddle point, as it has one negative eigenvalue and all other eigenvalues are positive.
			\begin{eqnarray}
				\label{eq:point5}
				A_5: \quad 	x=0, \qquad y=\pm \sqrt{\frac{\beta - 4 \sqrt{6}\xi z - 6 \beta \xi z^2}{\alpha+\beta}}, \nonumber\\ A=\sqrt{\frac{6 \alpha \xi z^2- 4 \sqrt{6}\xi z - \alpha}{\alpha+\beta}}, \qquad z=z.
			\end{eqnarray}
			This fixed point depends on the model parameters $(\alpha,\beta,\xi)$, while the variable $z$ remains arbitrary. However, it is subject to the constraint imposed by the coupling function $F(\phi)=1-\kappa^2 \xi \phi^2 \equiv 1- 6 \xi z^2$. To ensure the absence of fatal instabilities, $F(\phi)$ must remain strictly positive. This requirement leads to $6 \xi z^2 < 1$ which implies that $z < \frac{1}{\sqrt{6 \xi}}$. Substituting the parameter values listed in Tab.~\ref{tab:mean_value_para}, the dynamical variables asymptotically approach the values shown in Fig.~\ref{fig:dyn_stab_evo}. For the parameter values corresponding to Model~III, $(\alpha \sim 10^{-4}, \beta = 0.26, \xi = 0.12)$ with $z\sim10^{-3}$, the corresponding eigenvalues are
			\begin{equation}
				\label{eq:A5}
				\mu_1=-3.0,\qquad
				\mu_2=-2.4,\qquad
				\mu_3=-0.6,\qquad
				\mu_4=-10^{-6}.
			\end{equation}
			The fixed point $A_5$ represents a one-parameter family of equilibrium points, since the variable $z$ remains arbitrary. Consequently, one eigenvalue is expected to be exactly zero, corresponding to perturbations along the equilibrium curve. Therefore, the stability of $A_5$ should not be justified by requiring all eigenvalues to be negative. Instead, the correct interpretation is that the three transverse eigenvalues are negative, causing nearby trajectories to converge toward the equilibrium curve. Once the trajectories reach the curve, they remain on it, as shown in Fig. 9.} Furthermore, the phase portrait demonstrates that the trajectories originating from different regions of the phase space converge toward the critical point $A_5$, confirming its asymptotically stable attractor nature, see Fig.~\ref{fig:port}. A summary of all the critical points is provided in Tab.~\ref{tab:FX}.
		Corresponding to this case, $\frac{\dot{H}}{H^2}=0$, $\Omega_{\phi}=1$ and $w_{\rm eff}=w_{\phi}=-1$, implies
	\begin{eqnarray}
		\label{eq:a}
		a(t)&=&a_0 e^{H_0(t-t_0)}.
	\end{eqnarray}
	In order to derive the expression for $\phi(t)$, we will utilize the subsequent dimensionless variable.
	\begin{equation}
		x=\frac{\kappa \dot{\phi}}{\sqrt{6}H}
	\end{equation}
	At the stationary point where $x=0$, it follows that $\dot{\phi}=0$, leading to the conclusion that $\phi=\phi_0$ (constant). It is evident that both $\dot{H}=\dot{\phi}=0$, indicating that this point exhibits de-Sitter characteristics. {The phase-space analysis and the observational results are in close agreement, as both indicate that the de Sitter attractor ($A_5$) represents the late-time cosmological epoch. This conclusion is further supported by the evolution illustrated in Fig.~\ref{fig:evo_total}. }
	Let us examine the behavior of $F(\phi)=1-\kappa^2 \xi \phi^2 \equiv 1- 6 \xi z^2$. 
	\begin{enumerate}
		\item For the critical point $A_1$, we have $z=0$, which gives $F(\phi)=1$.
		\item For the critical point $A_2$, we obtain $z=\pm \frac{1}{\sqrt{6 \xi}}$, in this case, $F(\phi)=0$. 
					\item For the critical points $A_3$ and $A_4$, the values of $z$ are given by Eqs.~(\ref{eq:point3}) and ~(\ref{eq:point4}). In this case, we find $F(\phi) \simeq 1$ and 0.96, respectively, for the observationally preferred parameters.
			\item For the critical point $A_5$, the value of $z$ is arbitrary, see Eq.~(\ref{eq:point5}). To avoid fatal instabilities, $F(\phi)$ must be positive, which is possible when $z < \frac{1}{\sqrt{6 \xi}}$. In this case, we find $F(\phi) \simeq 1$ for $z=0.001$ and $\xi=0.12$.
	\end{enumerate}
	\begin{table}[tbp]
		\centering		
		\resizebox{\columnwidth}{!}{
			\begin{tabular}{c c c c c c c c c}
				\hline
				\hline
				Point  & $x$  & $y$ & $A$ & $z$  & Eigenvalues & Stability  & $\Omega_{\phi}$  & $w_{\rm eff}$\\
				\hline
				$A_1$ & 0 & 0 & 0 & 0 &  Eq. (\ref{eq:A1}) & Saddle & 0 & 0\\
				$A_2$ & 0 & 0 & 0 & $\pm \frac{1}{\sqrt{6 \xi}}$ & Eq. (\ref{eq:A2}) & Saddle & 1 & 1/3\\
				$A_3$ & 0 & $\pm \frac{2\sqrt{-2\xi+\sqrt{\xi(\alpha^2+4\xi)}}}{\alpha}$ & 0 & $\frac{2\sqrt{6}\xi \pm \sqrt{6\alpha^2 \xi+24 \xi^2}}{6 \alpha \xi}$ & Eq. (\ref{eq:A3}) & Non-hyperbolic & 1 & $-1$ \\
				$A_4$ & 0 & $0$ & $\frac{2\sqrt{2\xi+\sqrt{\xi(\beta^2+4\xi)}}}{\beta}$ & $\frac{-2\sqrt{6}\xi + \sqrt{6\beta^2 \xi+24 \xi^2}}{6 \beta \xi}$ & Eq. (\ref{eq:A4}) & Saddle & 0 & 0 \\
				$A_5$ & 0 & $\pm \sqrt{\frac{\beta - 4 \sqrt{6}\xi z - 6 \beta \xi z^2}{\alpha+\beta}}$ & $\sqrt{\frac{6 \alpha \xi z^2- 4 \sqrt{6}\xi z - \alpha}{\alpha+\beta}}$ & $z$ & Eq. (\ref{eq:A5}) & Stable & 1 & $-1$ \\
				\hline
				\hline
			\end{tabular}
		}		
		\caption{The table presents the stationary points, corresponding eigenvalues, and relevant cosmological parameters for the model under consideration. The $\Omega_{\phi}$  and $w_{\rm eff}$ are calculated for the observationally preferred values of the model parameters.}
		\label{tab:FX}
	\end{table} 
	
	\subsection{Cosmological Importance}
	\label{sec:CI}
	
	The critical points described above provide a road map for understanding the possible scenarios for the evolution of the universe. By analyzing their stability and cosmological implications, we can determine whether the universe is currently in an accelerating phase, a decelerating phase, or moving towards a stable attractor. Below, we shall break down the points in a more detailed manner, expanding on the differences in the cases and referring to the literature for comparison.

	\textbf{Point $A_1$ (Saddle Point):} The fixed point $A_1$ behaves as a saddle, which means it has a combination of stable and unstable directions. The point corresponds to a non-accelerating universe with $\Omega_{\phi}=0$, indicating that the scalar field is not contributing to the energy density at this point, and the total EoS $w_{\rm eff}=0$ signifies a universe that is neither accelerating nor decelerating in this regime. The saddle points often correspond to intermediate phases of the universe with energy components like matter dominating.

	\textbf{Point $A_2$ (Saddle Point):} This point also exhibits saddle behavior with two eigenvalues positive and the other two negative. This suggests that it is unstable in one direction while stable in others. This point corresponds to accelerating or non-accelerating phase which depends on the value of $\xi$: $\Omega_{\phi}=\xi$ and $w_{\rm eff}=-1+(1-\xi+8\xi^2)/(1-\xi+6\xi^2)$. The saddle point is often seen in models where different energy components (e.g., matter and dark energy) compete for dominance. The specific behavior of such point could vary based on the specific dynamics of the scalar field and matter content, but in general, saddle point like this is indicative of intermediate states in the evolution of universe.

			\textbf{Point $A_3$ (Non-hyperbolic Point):} The fixed point $A_3$ shows non-hyperbolic behavior as one of the eigenvalues is zero, see Eq.~\eqref{eq:A3}. At this point, the Universe undergoes accelerated expansion with $\Omega_{\phi}=1$, indicating complete domination by the scalar field, while $\Omega_{m}=0$ implies the absence of a matter component. The corresponding equations of state satisfy $w_{\rm eff}=w_{\phi}=-1$, which is equivalent to a cosmological constant.
			
			\textbf{Point $A_4$ (Saddle Point):} Similar to point $A_1$, this point also exhibits saddle behavior for the observationally favored values of the model parameters, indicating the presence of both stable and unstable directions in the phase space. At this point, $\Omega_{\phi}=0$, showing that the scalar field makes no contribution to the total energy density of the universe. In addition, the effective EoS $w_{\rm eff}=0$, corresponding to a non-accelerating cosmological phase. Such saddle points generally represent transient stages in the cosmic evolution, often associated with matter-dominated eras before the system evolves toward a stable late-time solution.
			
			\textbf{Point $A_5$ (Stable Attractor Solution):} This fixed point depends on the model parameters $(\alpha,\beta,\xi)$, while $z$ remains arbitrary. Substituting the best-fit parameter values listed in Tab.~\ref{tab:mean_value_para}, the dynamical variables defined in Eq.~\eqref{eq:var} asymptotically approach the values shown in Fig.~\ref{fig:dyn_stab_evo}. The analysis further shows that the critical point $A_5$ exists for all the models, with the coordinate $z$ approaching a small but non-zero value as shown in Fig.~\ref{fig:dyn_stab_evo}. Moreover, for the parameters constraint value of each model, the effective equation of state asymptotically approaches $w_{\rm eff}\rightarrow -1$, while the dark-energy density parameter tends to $\Omega_{\phi}\rightarrow1$. These results are fully consistent with the findings of the previous section, confirming that the cosmological evolution converges to a stable de Sitter phase.
			
			To further illustrate the attractor behavior, we plot the reduced phase space by fixing the planes $A=0.1$ and $z=0.001$, as shown in Fig.~\ref{fig:port}. The trajectories converge toward the fixed point $A_5$, confirming its stability. Moreover, this late-time de Sitter attractor is fully consistent with the observational analysis presented in Sec.~\ref{sec:RES}, demonstrating that the observationally preferred cosmological trajectories evolve toward the stable fixed point $A_5$.
			

	\section{Conclusion}
	\label{sec:conc}
	In this paper, we investigated the cosmological dynamics of the Q-SC-CDM model in the presence of a non-minimal coupling parameter $\xi$. We began our analysis with the well-known $F(\phi)R$ coupling and carried out a detailed study of the cosmological evolution. We found that, for lower values of the coupling parameter, $\xi\lesssim 0.2$, the model gives rise to a future slow-contraction phase, often associated with a Big Crunch scenario, depending on the combinations of potential parameters $(\alpha, \beta)$ whenever they take higher values, see Fig. \ref{fig:NMC1}. Such solutions, fail to produce a sufficiently prolonged matter-dominated epoch at high redshift, which is essential for explaining the observed Universe. At the same time, we identified regions of parameter space with $\xi \lesssim 0.2$ that do not exhibit a future slow-contraction phase whenever combination of potential parameters are remain small, $|\alpha \cdot \beta|<10^{-0.5} $, and instead generate an extended matter-dominated era while closely mimicking the $\Lambda$CDM cosmology. As the coupling is increased to $\xi \gtrsim 0.2$, the slow-contraction phase disappears, and the Hubble parameter exhibits a transient evolution before asymptotically approaching a constant value. In this regime, the effective equation of state approaches $w_{\rm eff}=-1$, leading to a stable de Sitter phase in the future, see Fig. \ref{fig:NMC2}.
	
	To investigate the impact of $\xi$ on cosmological observables, we confronted the model with a broad combination of datasets, including CC, DESI BAO measurements, and three independent Type Ia supernova compilations: Pantheon+, DESY5, and Union3. We classified the system into three scenarios: Model I and Model II, corresponding to fixed values $\xi=0.3$ and $\xi=0.5$, respectively, and Model III, where $\xi$ was allowed to vary within the range $[0.01,0.7]$. Bayesian model comparison shows that all considered Models are strongly disfavored by the data relative to $\Lambda$CDM. However, the preferred lower values of $\xi$ which is corresponding to Model III, yield an evolution that closely resembles $\Lambda$CDM, including an extended matter-dominated epoch and a stable de Sitter attractor at late times, without exhibiting a slow-contraction phase. This further supports the conclusion that solutions displaying a future slow-contraction phase are generally unable to reproduce a sufficiently long matter-dominated era. Our fixed point analysis independently confirms that the observationally favored Model III is stable at the background level and evolves toward a stable de Sitter state.
	
	Future work should extend the analysis of Model III to the perturbation level and confront it with a broader range of cosmological observations, including the full CMB dataset and large-scale structure measurements. Since the background analysis constrains $\xi$ to relatively small values, it will be particularly interesting to investigate its impact on the matter clustering amplitude $\sigma_8$. Moreover, alternative forms of the scalar-field potential could be explored beyond the double-exponential potential considered here. Since the present potential leads to dynamics that closely track $\Lambda$CDM, other theoretically motivated choices, such as axion-inspired cosine potentials or combinations of cosine and exponential potentials, may generate richer cosmological behavior and potentially yield improved Bayesian evidence relative to the standard cosmological model.

	\begin{acknowledgements}
		We thank the anonymous referee for the insightful comments and constructive suggestions, which have significantly improved the presentation and overall quality of the manuscript.
		This research was funded by the Science Committee of the Ministry of Science and Higher Education of the Republic of Kazakhstan (Grant No. AP23487178). SH acknowledges the support of National Natural Science Foundation of China under Grants No. W2433018 and No. 11675143, and the National Key Research and development Program of China under Grant No. 2020YFC2201503. MS acknowledges Integral University, Lucknow for financial support through Seed Money Grant 2024-2025 (Project Sanction No.: IUL/ICEIR/SMP/2024-04) and MCN: IU/R\&D/2025-MCN0003753. MS also thanks the Inter-University Centre for Astronomy and Astrophysics (IUCAA), Pune for the hospitality and facilities under the visiting associateship program where the work was completed. 
	\end{acknowledgements}
	
	\appendix
	
	\section{The variation of $\xi$ with the model parameters \ensuremath{\alpha} and \ensuremath{\beta}\label{appen:variation_of_xi_with_alpha_beta}}
	
	\begin{figure}[tbh]
		\resizebox{\columnwidth}{!}{\includegraphics{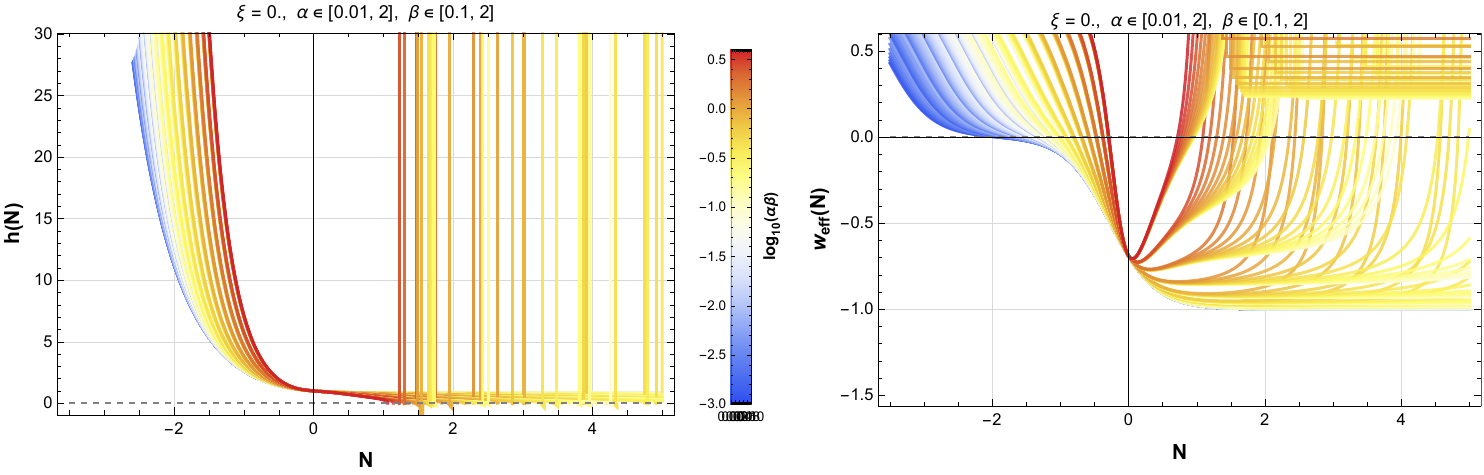}}
		\caption{Characteristics of the model for $\xi= 0$ varying for $\alpha \in [0.01,2]$ and $\beta \in [0.1, 2]$ by evolving equations Eq.~\eqref{x_prime}--Eq.~\eqref{om_prime} by fixing the initial conditions $x(0)= 10^{-4}, z(0) =0.04, A(0) = 0.1, \Omega_{0m} = 0.31$. }
		\label{fig:xi_0_alpha_beta}
	\end{figure}
	
	\begin{figure*}[tbh]
		\resizebox{\linewidth}{!}{\includegraphics{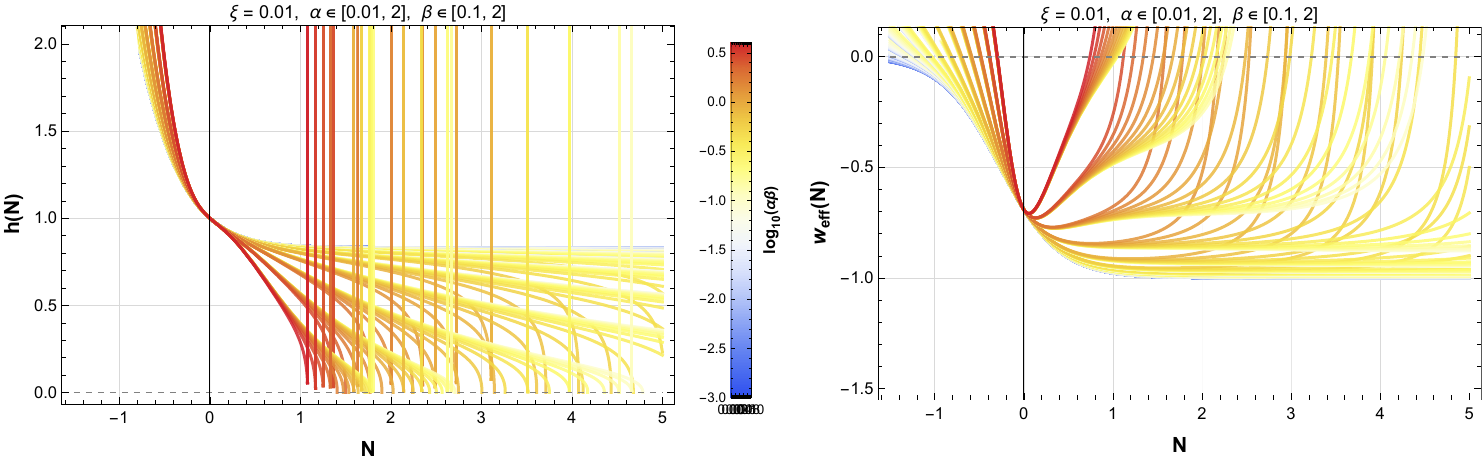}
			\includegraphics{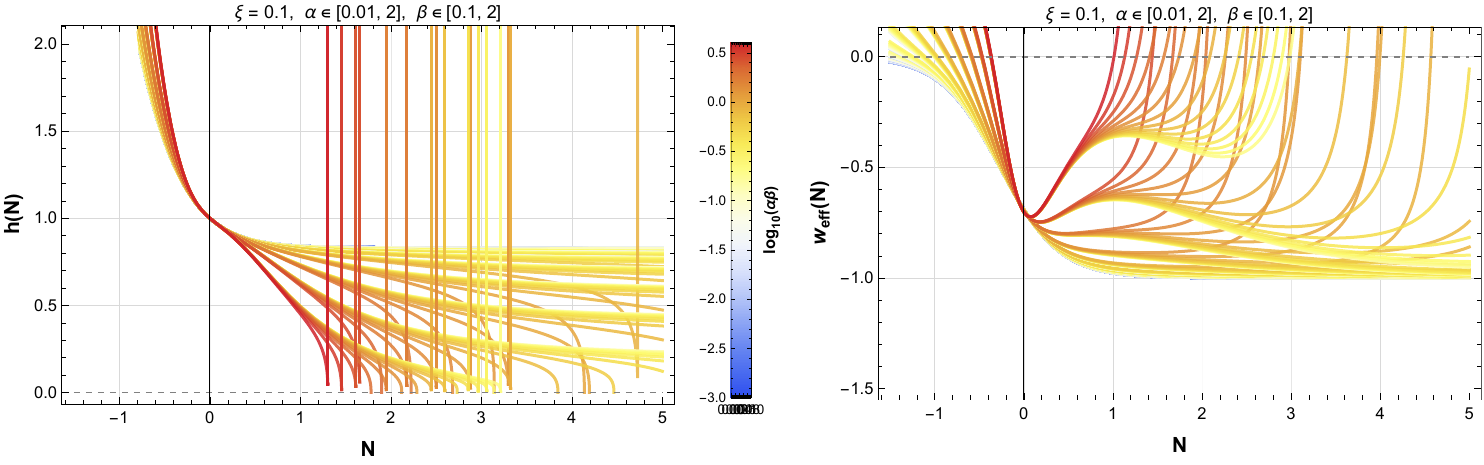}}
		
		\resizebox{\linewidth}{!}{\includegraphics{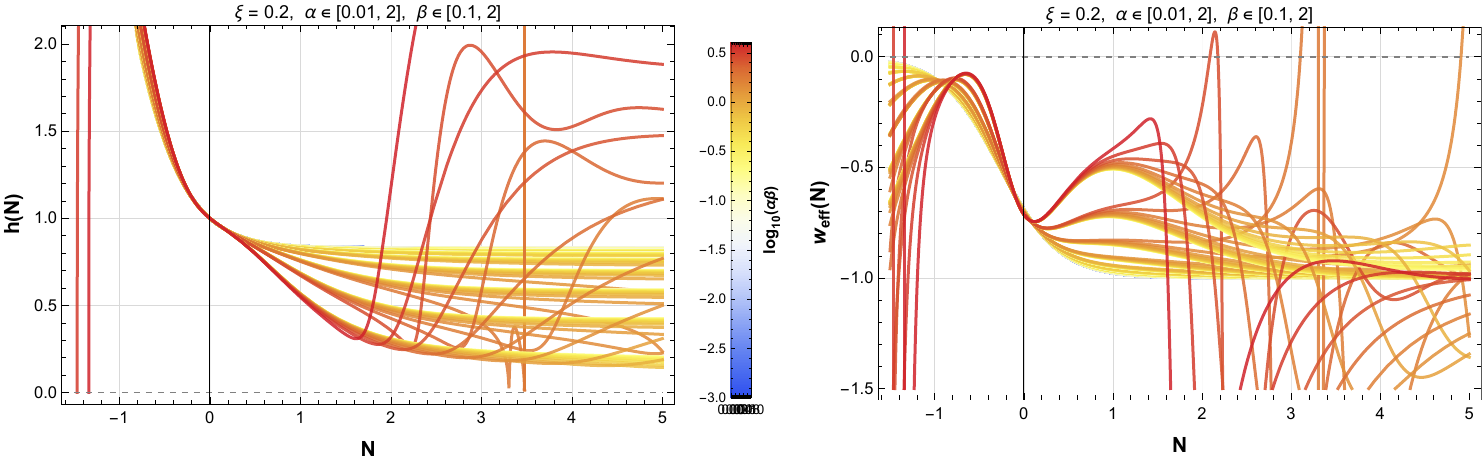}
			\includegraphics{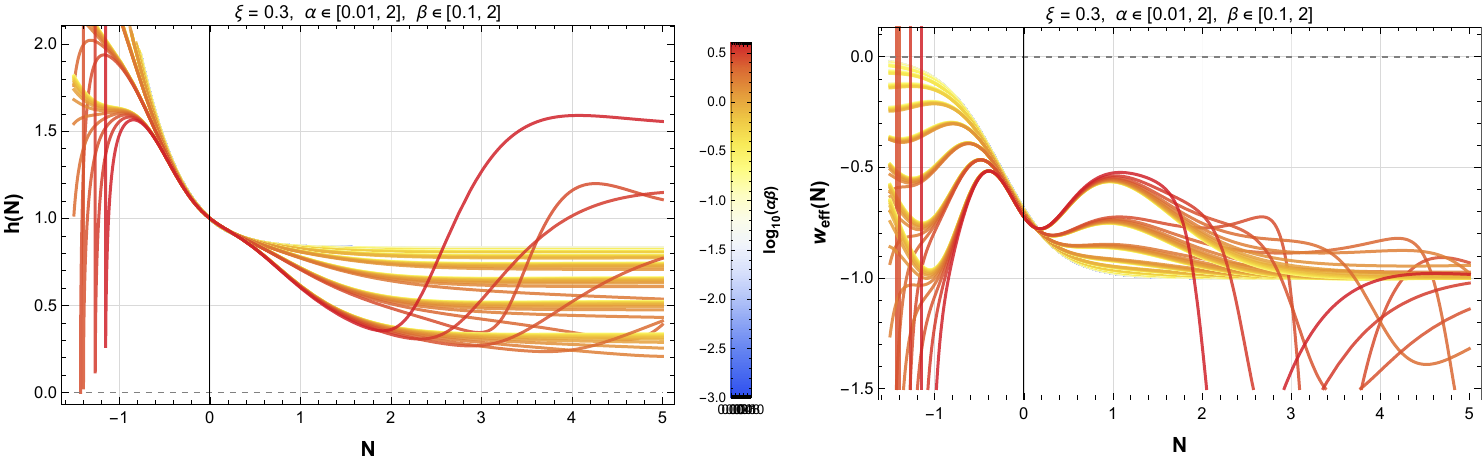}}
		
		\resizebox{\linewidth}{!}{\includegraphics{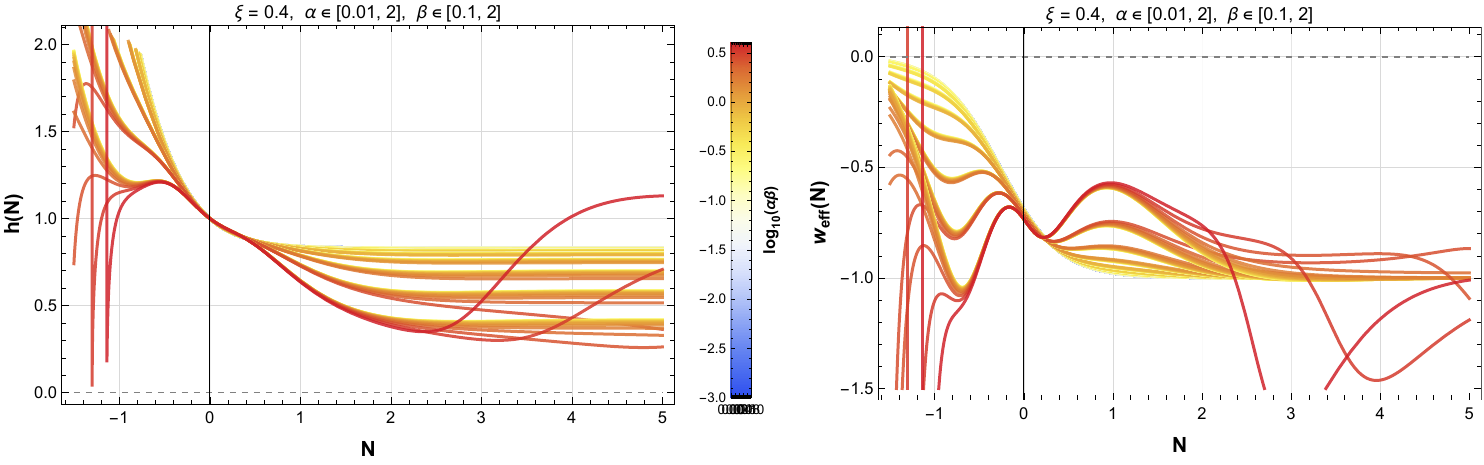}
			\includegraphics{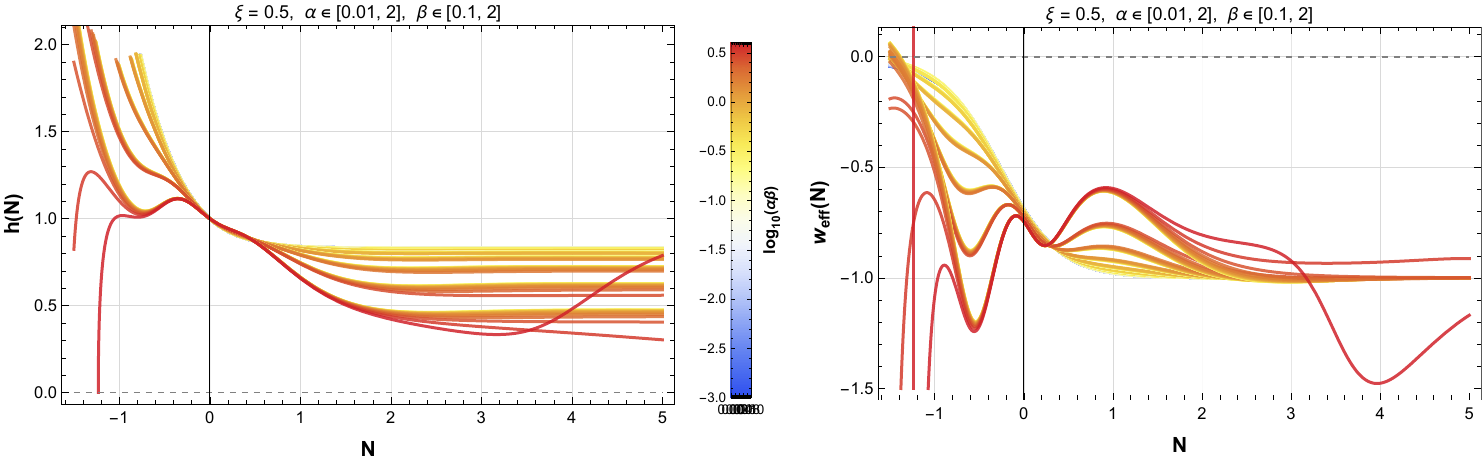}}
		
		\resizebox{\linewidth}{!}{\includegraphics{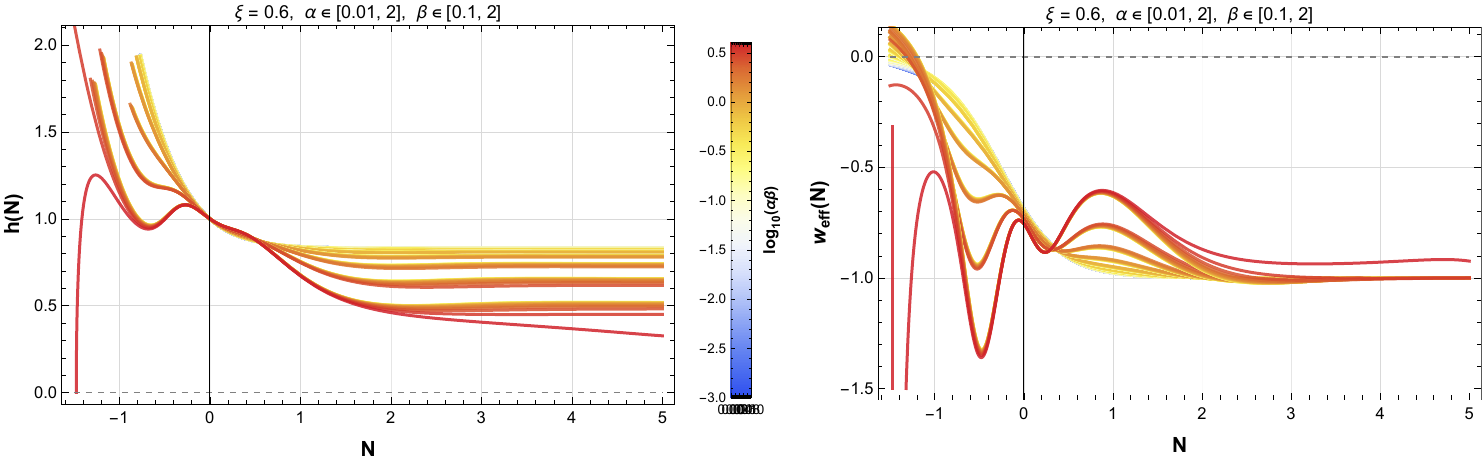}
			\includegraphics{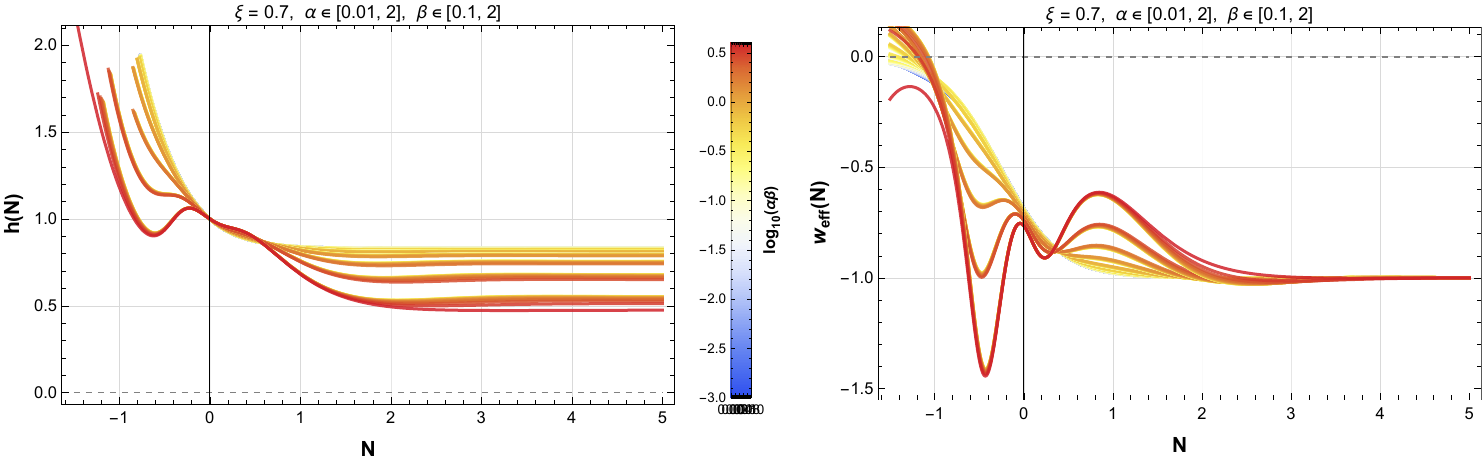}}
		
		\resizebox{\linewidth}{!}{\includegraphics{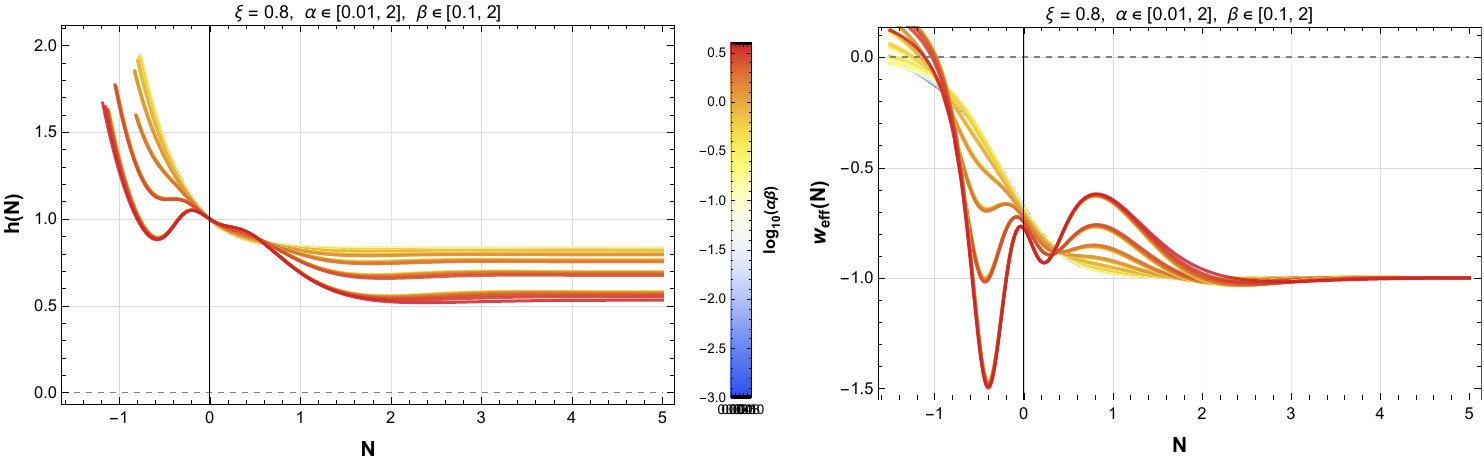}
			\includegraphics{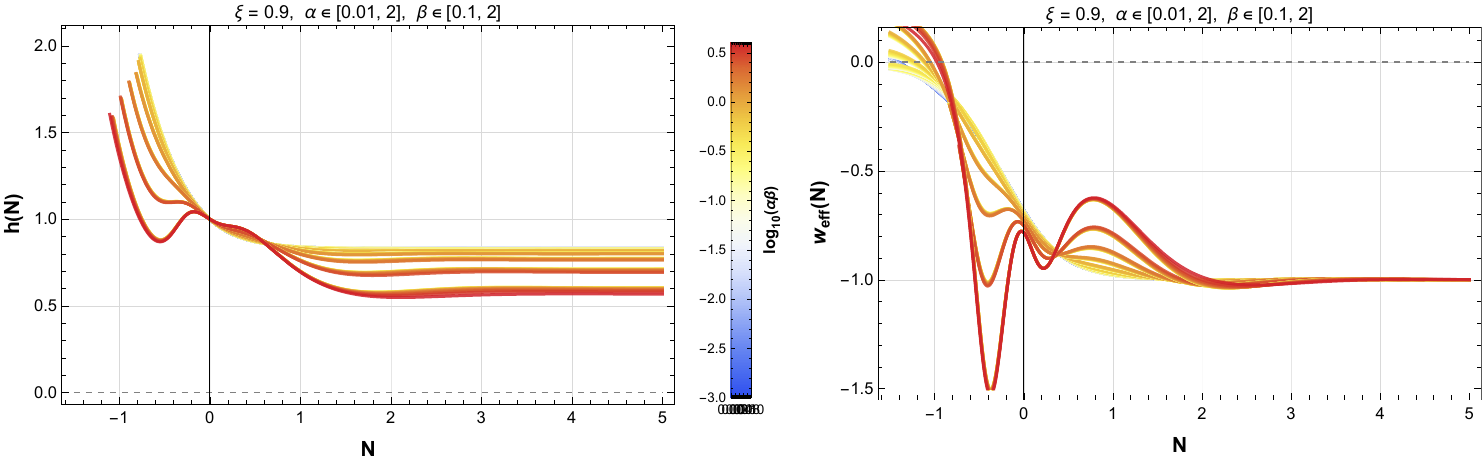}}
		\caption{Characteristics of the model for $0<\xi<1$ varying for $\alpha \in [0.01,2]$ and $\beta \in [0.1, 2]$ by evolving equations Eq.~\eqref{x_prime}--Eq.~\eqref{om_prime} by fixing the initial conditions $x(0)= 10^{-4}, z(0) =0.04, A(0) = 0.1, \Omega_{0m} = 0.31$. }
		\label{fig:xi_1_alpha_beta}
	\end{figure*}

\end{document}